\documentclass[preprint2]{aastex}
\shorttitle{Microlensing Analysis of SDSS J0924+0219}
\shortauthors{MacLeod et al.}
\usepackage{epstopdf}
\begin{document}
\title{A Consistent Picture Emerges: A Compact X-ray Continuum Emission Region in the Gravitationally Lensed Quasar SDSS J0924+0219}
\author{Chelsea L. MacLeod\altaffilmark{1,2}, Christopher W. Morgan\altaffilmark{1},
  A. Mosquera\altaffilmark{3}, C. S. Kochanek\altaffilmark{3,4}, M. Tewes\altaffilmark{5},
  F. Courbin\altaffilmark{6}, G. Meylan\altaffilmark{6},
  B. Chen\altaffilmark{7}, X. Dai\altaffilmark{7}, G. Chartas\altaffilmark{8}}
\altaffiltext{1}{Physics Department, United States Naval Academy, Annapolis, MD 21403}
\altaffiltext{2}{Institute for Astronomy, University of Edinburgh, Edinburgh EH9 3HJ, U.K.; cmacleod@roe.ac.uk}
\altaffiltext{3}{Department of Astronomy, The Ohio State University, Columbus, OH 43210}
\altaffiltext{4}{Center for Cosmology and Astroparticle Physics, The Ohio State University, Columbus, OH 43210}
\altaffiltext{5}{Argelander-Institut f\"ur Astronomie, Auf dem H\"ugel 71, 53121 Bonn, Germany}
\altaffiltext{6}{Laboratoire d'astrophysique, Ecole Polytechnique F\'ed\'erale de Lausanne (EPFL), Observatoire de Sauverny, 1290 Versoix, Switzerland}
\altaffiltext{7}{Homer L. Dodge Department of Physics and Astronomy, The University of Oklahoma, Norman, OK 73019}
\altaffiltext{8}{Department of Physics and Astronomy, College of Charleston, SC 29424}

\begin{abstract}
  We analyze the optical, UV, and X-ray microlensing variability of the lensed quasar SDSS J0924+0219 using six epochs of $Chandra$ data in two energy bands (spanning 0.4--8.0~keV, or 1--20~keV in the quasar rest frame), 10 epochs of F275W (rest-frame 1089\AA) Hubble Space Telescope data, and high-cadence $R$-band (rest-frame 2770\AA) monitoring spanning eleven years.  Our joint analysis provides robust constraints on the extent of the X-ray continuum emission region and the projected area of the accretion disk. The best-fit half-light radius of the soft X-ray continuum emission region is between $5\times 10^{13}$ and $10^{15}$~cm, and we find an upper limit of $10^{15}$~cm for the hard X-rays.  The best-fit soft-band size is about 13 times smaller than the optical size, and roughly $7 GM_{BH}/c^2$ for a $2.8\times 10^8M_{\odot}$ black hole, similar to the results for other systems.  We find that the UV emitting region falls in between the optical and X-ray emitting regions at $10^{14}$~cm~$< r_{1/2,{\rm UV}} < 3\times 10^{15}$~cm. Finally, the optical size is significantly larger, by 1.5$\sigma$, than the theoretical thin-disk estimate based on the observed, magnification-corrected $I$-band flux, suggesting a shallower temperature profile than expected for a standard disk.\\  
\end{abstract}

\maketitle

\section{Introduction}

Quasars are now known to play a major role in galaxy formation and cosmology through their effects on star formation and ionization \citep[e.g.,][]{tey11}. However, the detailed structure of the central engine has remained unclear, as well as the relationships between emission at different wavelengths. The main difficulty is that the luminous source is so compact, requiring sub-microarcsecond resolution for direct imaging. Studies have made use of the time domain to probe their inner structure, such as placing upper limits on the variable source size using characteristic timescale arguments \citep[e.g.,][]{smi63, vau03} or using reverberation mapping to probe the broad line region and calibrate black hole masses \citep[e.g.,][]{pet92}. Reverberation lags in highly variable AGN have been used to explore the continuum emission \citep{sha14} and the X-ray generation process \cite[e.g.,][]{cac13}. Variations in the profiles of Fe fluorescence lines can also be used to constrain the inner and outer radii of the Fe line reflection component of the accretion disk \citep[e.g.,][]{fab89,lao91}. Microlensing of gravitational lensed quasars provide another way to ``zoom in'' on the most compact regions in AGN and dissect their structures. 

Microlensing in strongly lensed quasars can be used to measure many physical properties of the quasar continuum source and the lens galaxy.  Since quasar microlensing was first described by \citet{cha79} and detected by \citet{cor91}, studies have successfully isolated its signal by analyzing observations of lensed quasars over time \citep[e.g.,][]{mor10,mos13}, wavelength \citep[e.g.,][]{bat08,mun11}, or both \citep[e.g.,][]{poi08,eig08,eig08a}.   Microlensing by stars in the lens galaxy can (de)magnify individual images in a multiply-imaged system,  causing the ratios between the image fluxes to be significantly different from the predictions of strong lens models fit to the observed macroscopic properties of the lens.  Due to relative motions between the source, lens, stars in the lens, and observer, these magnifications change with time, leading to uncorrelated variability in the light curves of lensed quasars.

The analysis of quasar flux ratio anomalies in single- or few-epoch multi-wavelength observations is a relatively inexpensive means of estimating the quasar continuum source size \citep[e.g.,][]{poo07,bat08,flo09,med11,mos11a}. While insensitive to the mass function \citep[e.g.,][]{lew95}, this requires assuming the mass of an average star in the lens galaxy $\langle M \rangle$ \citep{wam92} and that the macro-model lens magnifications and differential extinctions are well understood.  The choice of a linear or logarithmic prior on the source size also tends to significantly affect the resulting size estimates \citep{mos11a}, and the estimates can be further complicated for lenses with significant time delays, where the flux ratios can be contaminated by the intrinsic variability of the quasar. Moreover, microlensing magnification patterns are complex and frequently divided into ``active'' regions of many overlapping caustics and ``passive'' valleys with few caustics, and single-epoch data provide little constraint on the location of the source relative to these patterns. Monitoring lenses over time addresses these problems \citep{koc04,koc06,goi08} while simultaneously permitting time delay measurements \citep[e.g., recent measurements by][]{tew12,eul13,rat13}. In particular, the Bayesian Monte Carlo method of \citet{koc04} enables the extraction of probability densities for the physical parameters of interest (e.g., source size, $\langle M \rangle$, etc.) by modeling the microlensing variability using a Monte Carlo simulation of the observed light curves. 

A common conclusion of essentially all microlensing studies is that there is a systematic discrepancy between microlensing size measurements and theoretical thin disk sizes \citep{sha73} or the size estimated from the quasar's observed luminosity \citep[e.g.,][]{poo06,bla11,mos11b,mun11,hai13}. The larger disk sizes inferred from microlensing suggest a shallower temperature profile for the disk, as might be found if irradiation of the outer disk by the inner disk is important \citep[see][]{mor10}. It does not appear to be explainable by emission line contamination or scattering \citep[see][]{dai10,mor10}. The temperature profile can be determined by measuring the dependence of the disk size on wavelength \citep[e.g.,][]{poi08,eig08,bla14,jim14}, but there is a large scatter in the best-fit dependence among studies.  

The X-ray emission regions are much more strongly microlensed than the optical \citep[e.g.,][]{mor01,cha02,dai03,bla06,poo07}, indicating that the X-ray source must be more compact. In fact, the analyses of microlensing variability for six quasars \citep{mor08,mor12,cha12,mos13,bla13,bla14} from our \emph{Chandra} monitoring program have consistently shown that the projection of the X-ray continuum emission region is confined to a region close to the inner edge of the accretion disk. The energy structure of the X-ray continuum source can be explored by considering the hard and soft X-ray bands separately, and these studies have presented mixed evidence for an energy-dependent X-ray size. For example, while \citet{che11} found a more compact hard X-ray source in Q2237+0305, \citet{cha12} found that the softest X-rays were microlensed more strongly than the hardest X-rays in image D by a factor 1.3 in RXJ1131$-$1231. Meanwhile, \citet{bla14} found no difference between the hard and soft X-ray sizes in HE 0435$-$1223.

The quadruply-imaged lensed quasar SDSS J0924+0219 \citep{ina03}, at a source redshift of 1.524, shows the most anomalous optical flux ratio of any four-image lens, and it has been unclear whether the anomaly is due to microlensing by stars \citep{pee04,kee06,mor06,flo09} or to perturbations to the lens potential by dark matter substructure \citep{eig06,fau11,bat11,koc04b}.  A comparison of the broad emission line flux ratios to the continuum flux ratios suggests that both substructure and microlensing must significantly contribute to the anomalous A/D flux ratio \citep{kee06}. Furthermore, variability in the X-ray continuum flux ratios \citep{poo12,che12} also indicates that microlensing must be present in this system.  \citet{mor06} analyzed the microlensing using optical monitoring data from the SMARTS telescope and found that the accretion disk radius as measured from microlensing is larger than, but marginally consistent with, that inferred from thin disk theory and the magnification-corrected $I$-band flux.   Later, \citet{flo09} found the size-wavelength power-law index $R_{\lambda}\propto \lambda^{\zeta}$ to be less than $\zeta<1.34$ for SDSS J0924+0219 at 95\% confidence, challenging the standard thin disk model (where $\zeta=4/3$), although such single-epoch studies generally yield prior-dependent results \citep{mos11a}.  Assuming that only microlensing is responsible for the observed flux anomaly, \citet{pee04} and \citet{mor06} predicted a substantial brightening of the faint saddle point image D on a timescale of roughly one decade. 

Here, we extend the microlensing analysis of \citet{mor06} using recent optical, ultraviolet (UV), and X-ray monitoring observations of SDSS J0924+0219. While image D has remained roughly an order of magnitude fainter than image A, allowing for the possibility of millilensing by dark matter substructure, our light curve-based microlensing analysis remains robust.  In Section~\ref{obs}, we describe the $R$-band monitoring data obtained by the SMARTS and Euler telescopes, the UV data from the {\it Hubble Space Telescope (HST)}, and the X-ray data from  \emph{Chandra}.  In Section~\ref{mod}, we describe our analysis technique and present our results from a joint Bayesian Monte Carlo microlensing analysis. In Section~\ref{disc}, we discuss our results. Where needed, we adopt a $\Omega_{M}=0.3, \Omega_{\Lambda}=0.7$ flat cosmology.

\section{Observations}
\label{obs}
\subsection{Optical Monitoring}

\begin{figure}[t!]
\centering  
\includegraphics[scale=.35,angle=270,trim=0cm 0cm 0cm 4cm,clip=true]{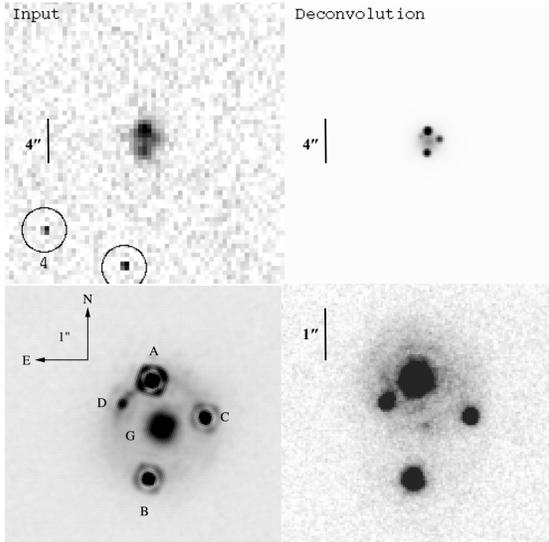}
\caption{\footnotesize SMARTS images for SDSS J0924+0219 (top two panels), showing the raw data (left) and the deconvolution (right) from May 1, 2004. Bottom-left: \emph{HST H}-band image. Bottom-right:  stacked \emph{HST} UV image (ten epochs over two years). }
\label{fig:data}
\end{figure}

\begin{figure*}[t!]
\centering  
\includegraphics[scale=.6]{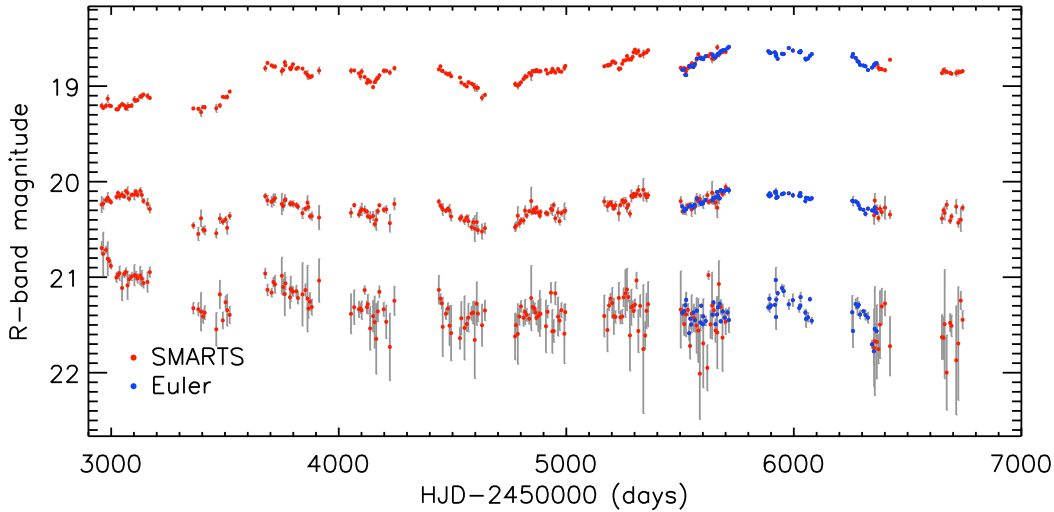}
\caption{\footnotesize $R$-band light curves for images A+D summed (top curve), B (middle curve), and C (bottom curve) in SDSS J0924+0219. There are 191 epochs of SMARTS data, and 60 epochs of Euler data (beginning on day $HJD-2450000 =5510$ in 2010 and ending on $HJD-2450000 =6364$). }
\label{fig:LC}
\end{figure*}

We monitored SDSS J0924+0219 for ten seasons (2003--2014) in the $R$ band using the SMARTS 1.3 m telescope with the ANDICAM optical/infrared camera \citep{dep03}\footnote{http://www.astronomy.ohio-state.edu/ANDICAM/}, and for three seasons (2010--2013) using the 1.2 m Euler Swiss Telescope as a part of the COSMOGRAIL project\footnote{http://www.cosmograil.org/}.  In the rest frame of the quasar at $z=1.524$, the effective wavelength of the $R$ band corresponds to 2770\AA. To obtain light curves of the quasar images, we use the photometric pipeline described in \cite{tew12}, which is based on the deconvolution algorithm of \citet{mag98}. We fit four point sources plus a Sersic profile to the lens components using astrometry from \citet{eig06}. Unfortunately, we were unable to reliably separate the flux of image D from that of image A, so we will fit the combined flux for images A and D with our Monte Carlo models.  A sample SMARTS image and its deconvolution are shown in Figure~\ref{fig:data}, along with the $HST$ $H$-band and UV images for reference.  We adjust the SMARTS and Euler light curves for images A+D, B, and C to minimize their dispersion in the overlapping observing seasons (2010/2011 and 2013).  This adjustment compensates for any small color terms introduced by differences in the filters and quantum efficiency curves of the CCDs, and it corrects for any PSF variability caused by minor differences in the flatness of each detector.  Our final optical light curves are shown in Figure~\ref{fig:LC} and provided in Table~\ref{tab:LC} (each ``season'' corresponds to a grouping of data points).  The average seeing was 1\farcs3, with a range of 0\farcs9 to 2\farcs0. The absolute photometric calibration of these light curves is obtained using SDSS DR9 $r$-band photometry of field stars. We estimate the zero-point uncertainty to be 0.07~mag. This calibration uncertainty is not included in the errors bars in Figure~\ref{fig:LC} or Table~\ref{tab:LC}, as the absolute calibration has no effect on our further analysis which only uses magnitude differences. 

\begin{deluxetable}{c c c c c}
  \tablecolumns{5}
  \tabletypesize{\footnotesize}
  \tablewidth{0pt}
  \tablecaption{ SDSS J0924+0219 Optical Light Curves \label{tab:LC}}
  \tablehead{HJD $-$ 2,450,000 & QSO A+D & QSO B & QSO C & Source}
  \startdata
  5543.806 &  $18.786 \pm 0.011$&$  20.243 \pm 0.027 $&$  21.720 \pm 0.136 $& SMARTS\\
  5544.828 &  $18.793 \pm 0.003$&$  20.255 \pm 0.007 $&$  21.434 \pm 0.018 $&  Euler\\
  5548.821 &  $18.794 \pm 0.003$&$  20.258 \pm 0.007 $&$  21.495 \pm 0.045 $&  Euler\\
  5550.755 &  $18.763 \pm 0.022$&$  20.291 \pm 0.073 $&$  21.382 \pm 0.197 $& SMARTS\\
  5558.831 &  $18.799 \pm 0.004$&$  20.271 \pm 0.012 $&$  21.370 \pm 0.024 $&  Euler\\
  5559.694 &  $18.782 \pm 0.016$&$  20.253 \pm 0.029 $&$  21.425 \pm 0.143 $& SMARTS\\
  5564.789 &  $18.765 \pm 0.003$&$  20.242 \pm 0.009 $&$  21.451 \pm 0.054 $&  Euler\\
  5565.761 &  $18.773 \pm 0.022$&$  20.262 \pm 0.077 $&$  21.446 \pm 0.130 $& SMARTS\\
  5572.758 &  $18.723 \pm 0.010$&$  20.170 \pm 0.047 $&$  21.506 \pm 0.291 $& SMARTS\\
(5573.771) &  $18.711 \pm 0.005$&$  20.179 \pm 0.007 $&$  21.350 \pm 0.018 $&  Euler\\
  \tableline 
  \enddata
  \tablecomments{\footnotesize{ Light curves are in calibrated $R$-band magnitudes.  Epochs in parentheses were rejected as outliers (see Section~\ref{sec:optonly}). Only a portion of this table is shown here to demonstrate its form and content. }}
\end{deluxetable}

\subsection{UV Data}
\begin{figure*}[ht!]
\centerline{
\includegraphics[scale=.35]{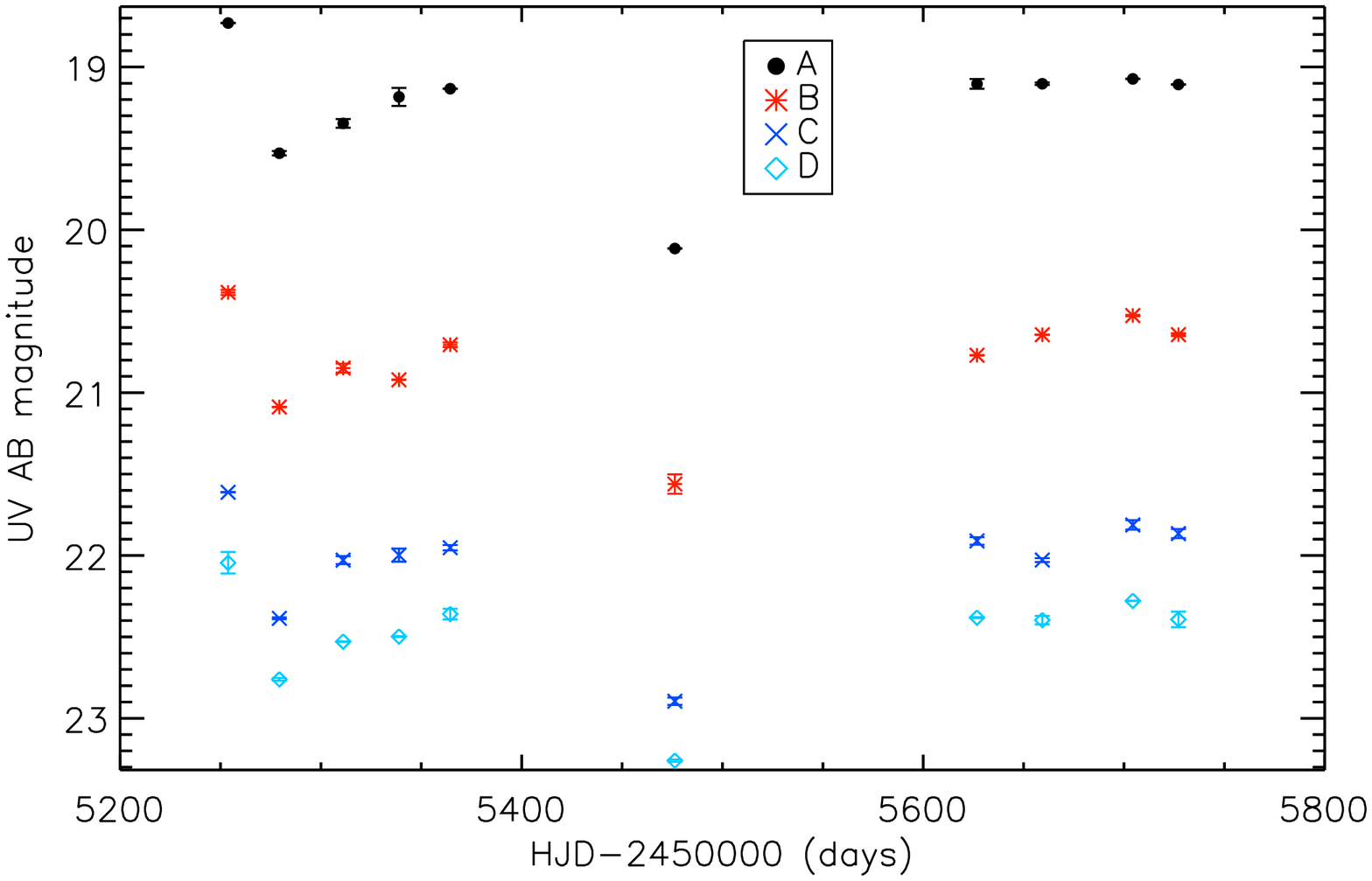}
\includegraphics[scale=.35]{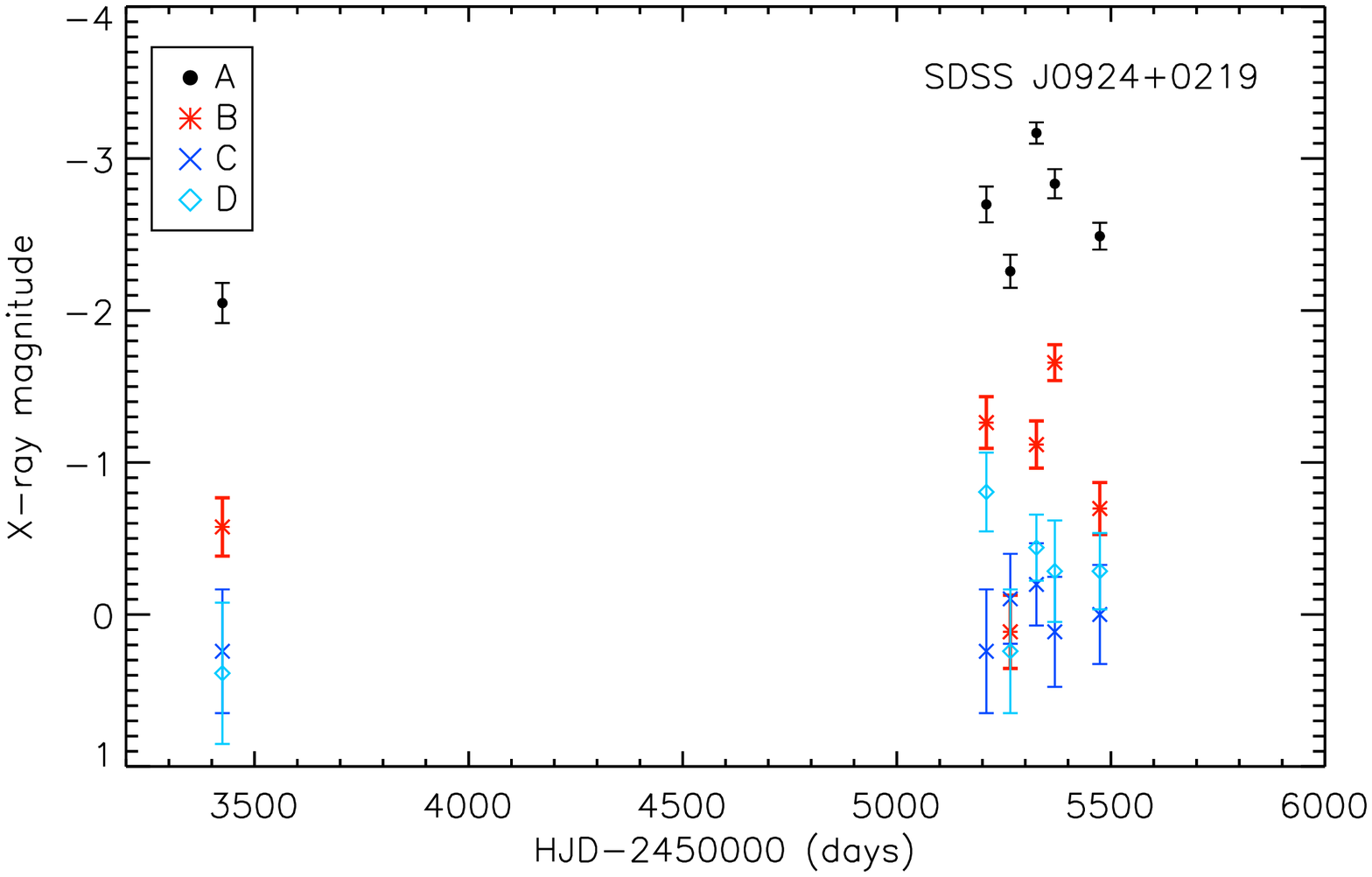}}
\centerline{
\includegraphics[scale=.35]{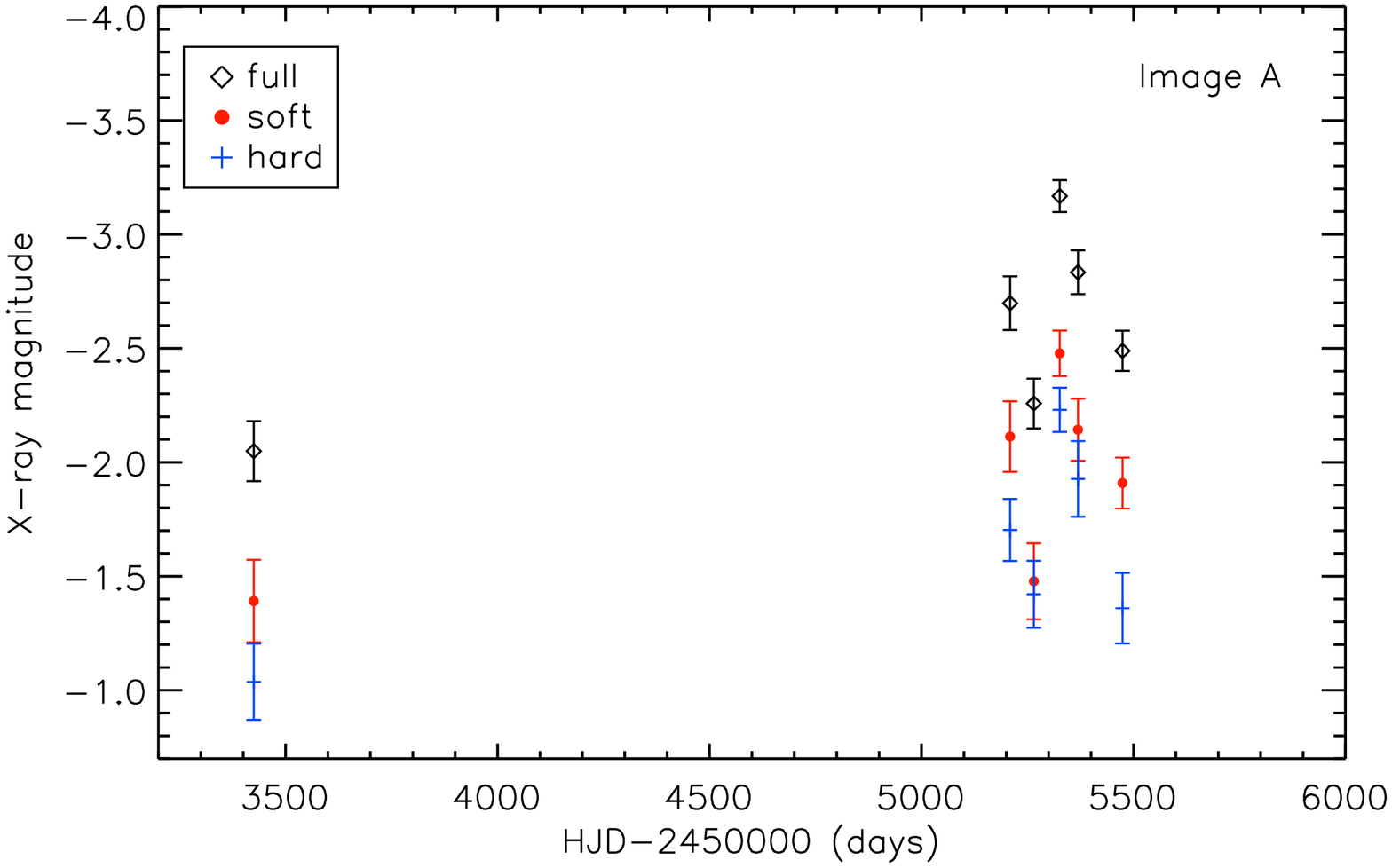}
\includegraphics[scale=.35]{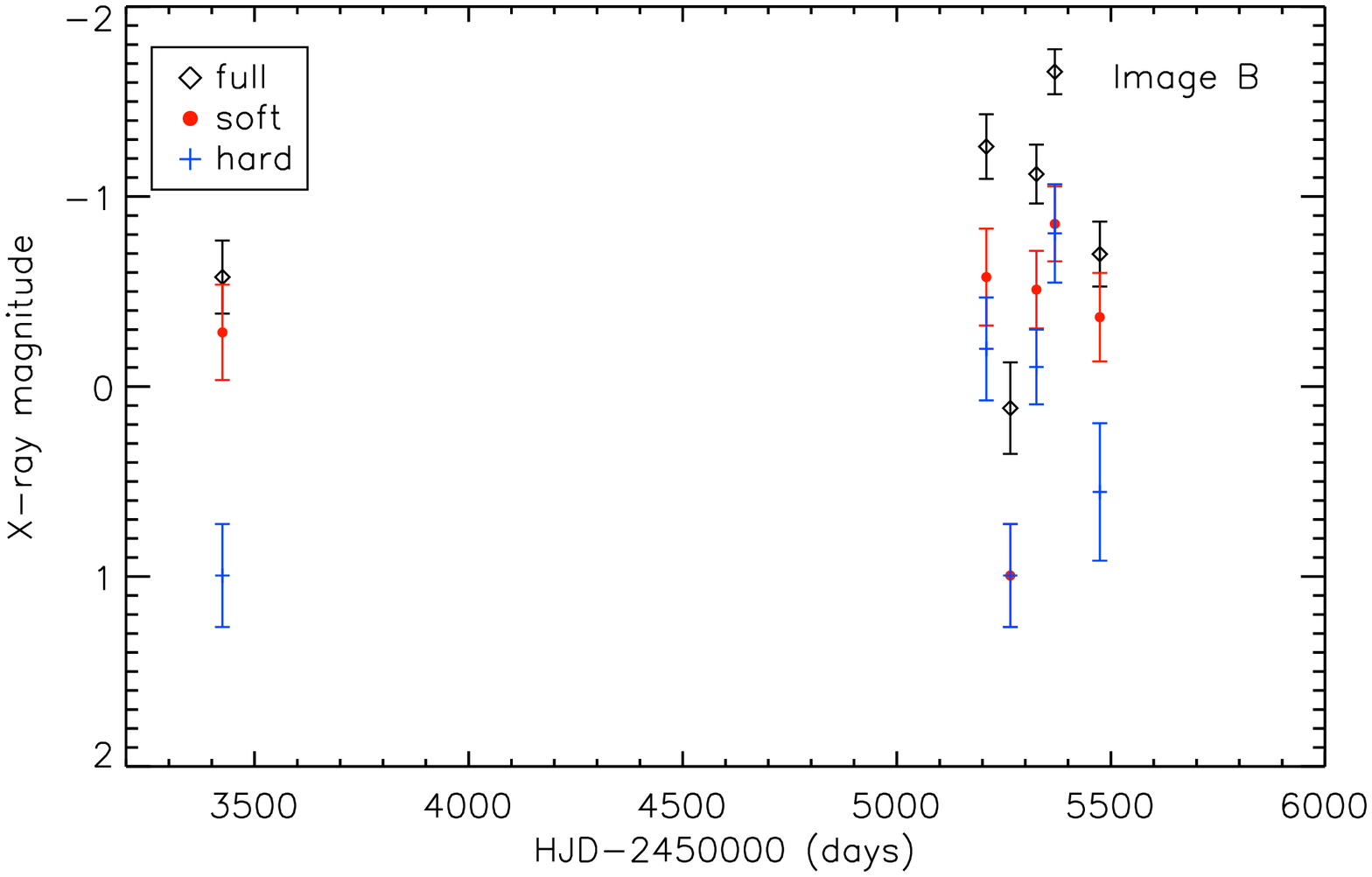}}
\centerline{
\includegraphics[scale=.35]{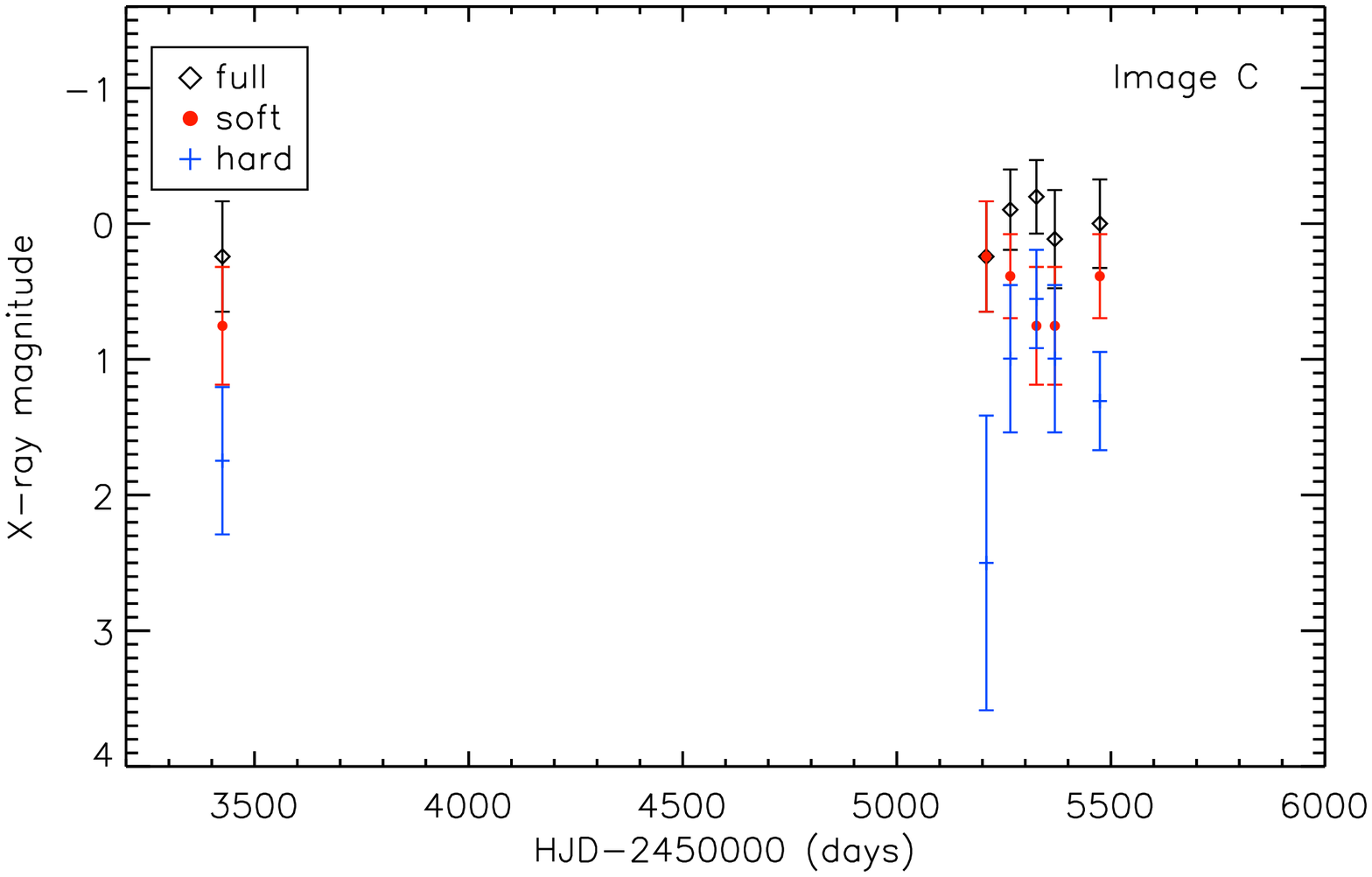}
\includegraphics[scale=.35]{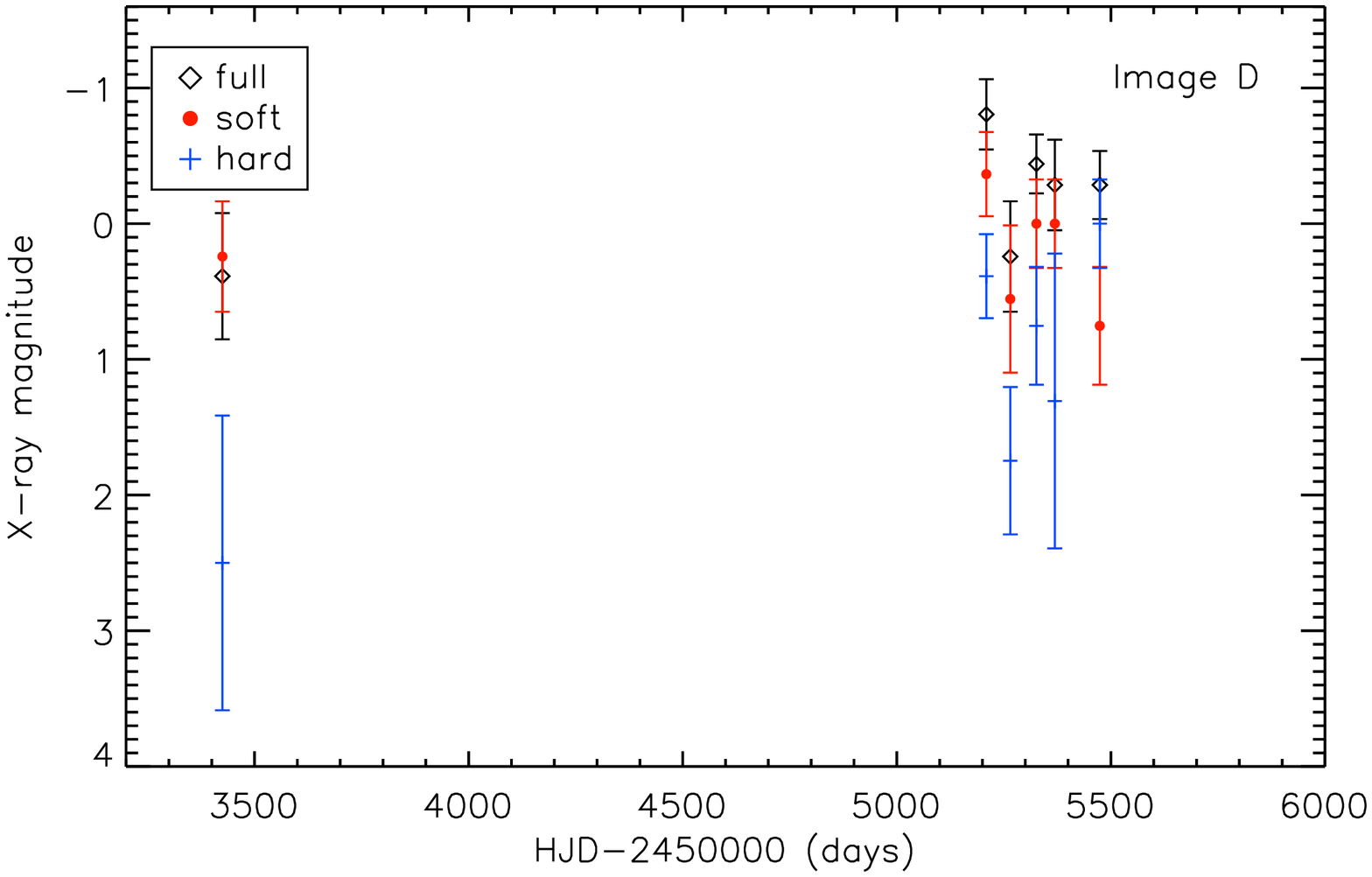}}
\caption{\footnotesize Top-left panel:  \emph{HST} UV light curves in AB magnitudes for the four quasar images in SDSS J0924+0219. Top-right panel: full-band X-ray light curves for SDSS J0924+0219 from \citet{che12}. The remaining panels show the X-ray data split into hard and soft bands for each image individually.}
\label{fig:LCx}
\end{figure*}

 SDSS J0924+0219 was observed at 10 epochs between 2009 August and 2011 August using the Wide Field Camera 3 (WFC3) on board the \emph{Hubble Space Telescope (HST)}. The observations used the F275W filter in the UVIS channel, providing flux ratios at 2750\AA\ (1089\AA\ in the source rest frame). The details of the reduction of these observations are presented in \citet{bla14}.  We use \emph{imfitfits} \citep{mcl98,leh00} to perform PSF photometry and assign error bars that correspond to the difference between the flux ratio estimates from PSF and aperture photometry.  We then use the PHOTFLAM, PHOTPLAM, and PHOTZPT header keywords to convert the instrumental magnitudes to the AB system (in this filter, the offset from ST to AB magnitudes is $m_{\rm AB} - m_{\rm ST} = 1.532$). We provide the calibrated photometry in Table~\ref{tab:uv}. We are unable to provide photometry for nearby comparison objects due to the lack of other UV-bright sources in the small \emph{HST} field. Figure~\ref{fig:LCx} shows the final UV light curves. Note that while images A and D are merged in the optical data, they remain separate in the UV light curve analysis, with a mean flux ratio $f_A/f_D=20\pm1$.

As the lens galaxy is expected to be faint in the UV, we searched the stacked UV image for the fifth, central image of the quasar which is expected from the lens model. To test for a central image, we used PSF fitting to carefully subtract the wings of the four bright quasar images. The PSF model was constructed from the exposure-weighted sum of oversampled and appropriately rotated Tiny Tim \citep{kri11} model PSFs for each of the 10 epochs. We fixed the relative positions of the five quasar images using the positions of the images and the lens galaxy from the CfA-Arizona Space Telescope Lens Survey (CASTLES\footnote{http://www.cfa.harvard.edu/castles/}), allowing the overall position offset and the five normalizations to vary. The residuals suggest the presence of a fifth image, but the detectability is unstable under changes in the PSF model. We can set an upper limit on its flux relative to image B of $f_5/f_B < 0.006$.  Such a detection, or limit, is not in conflict with the expectations from models of early-type galaxies \citep{kee05} and could be further weakened by any dust at the core of the lens galaxy.  Of course, a detection might also be problematic because it could also arise from nuclear star formation or weak AGN activity in the lens galaxy, which would be difficult to distinguish from the source quasar emission.

\begin{deluxetable}{c c c c c}
\tablecolumns{5}
\tablewidth{0pt}
\tabletypesize{\footnotesize}
\tablecaption{ SDSS J0924+0219 Ultraviolet (F275W) Light Curves \label{tab:uv}}
\tablehead{HJD $-$ 2,450,000 & QSO A & QSO B & QSO C & QSO D }
\startdata
 5253.8 & $18.730 \pm  0.001$ & $20.385  \pm 0.016$ & $21.612 \pm 0.001$  &  $22.045 \pm 0.066$  \\
 5279.3 & $19.530 \pm  0.013$ & $21.088  \pm 0.001$ & $22.386 \pm 0.005$  &  $22.760 \pm 0.008$  \\
 5311.1 & $19.346 \pm  0.026$ & $20.849  \pm 0.028$ & $22.028 \pm 0.025$  &  $22.529 \pm 0.001$  \\
 5338.9 & $19.184 \pm  0.055$ & $20.921  \pm 0.001$ & $21.998 \pm 0.040$  &  $22.498 \pm 0.002$  \\
 5364.4 & $19.134 \pm  0.001$ & $20.706  \pm 0.015$ & $21.952 \pm 0.016$  &  $22.360 \pm 0.032$  \\
 5476.3 & $20.115 \pm  0.001$ & $21.561  \pm 0.060$ & $22.895 \pm 0.023$  &  $23.260 \pm 0.007$  \\
 5626.8 & $19.104 \pm  0.030$ & $20.770  \pm 0.001$ & $21.911 \pm 0.023$  &  $22.382 \pm 0.002$  \\
 5659.3 & $19.103 \pm  0.008$ & $20.644  \pm 0.001$ & $22.028 \pm 0.012$  &  $22.397 \pm 0.026$  \\
 5704.4 & $19.073 \pm  0.001$ & $20.526  \pm 0.004$ & $21.812 \pm 0.029$  &  $22.279 \pm 0.001$  \\
 5727.1 & $19.108 \pm  0.001$ & $20.644  \pm 0.009$ & $21.866 \pm 0.028$  &  $22.393 \pm 0.048$  \\
\tableline
\enddata
\tablecomments{\footnotesize{  Light curves are in AB magnitudes. In this filter, the offset from ST to AB magnitudes is $m_{\rm AB} - m_{\rm ST} = 1.532$.  }}
\end{deluxetable}

\subsection{X-ray Data}

We monitored SDSS J0924+0219 in X-rays using the ACIS imaging spectrometer \citep{gar03} on the \emph{Chandra X-Ray Observatory} for six epochs between 2005 February 24 and 2010 October 5. 
 These observations were a component of a larger \emph{Chandra} Cycle 11 monitoring program, the details of which are published in \citet{che12}. As with the UV data, images A and D remain separate in the X-ray light curve analysis. The full-band data are shown in Figure~\ref{fig:LCx}. Note that while we only show the relative X-ray magnitudes, the calibrated fluxes are available in \citet{che12}. We also use magnitudes rather than simply logarithms of the flux so that the X-ray flux ratios are on the same scale as the optical/UV flux ratios.    We divided the full observed frame X-ray spectral range (0.4-8.0 keV) into soft (0.4-1.3 keV) and hard (1.3-8.0 keV) energy bands. This corresponds to a full-band range of 1--20~keV in the quasar rest frame. We analyzed the soft and hard X-ray variability separately to test for consistency and to explore any possible energy structure in the X-ray continuum source.

\begin{figure*}[ht!]
\centering
\includegraphics[scale=.35]{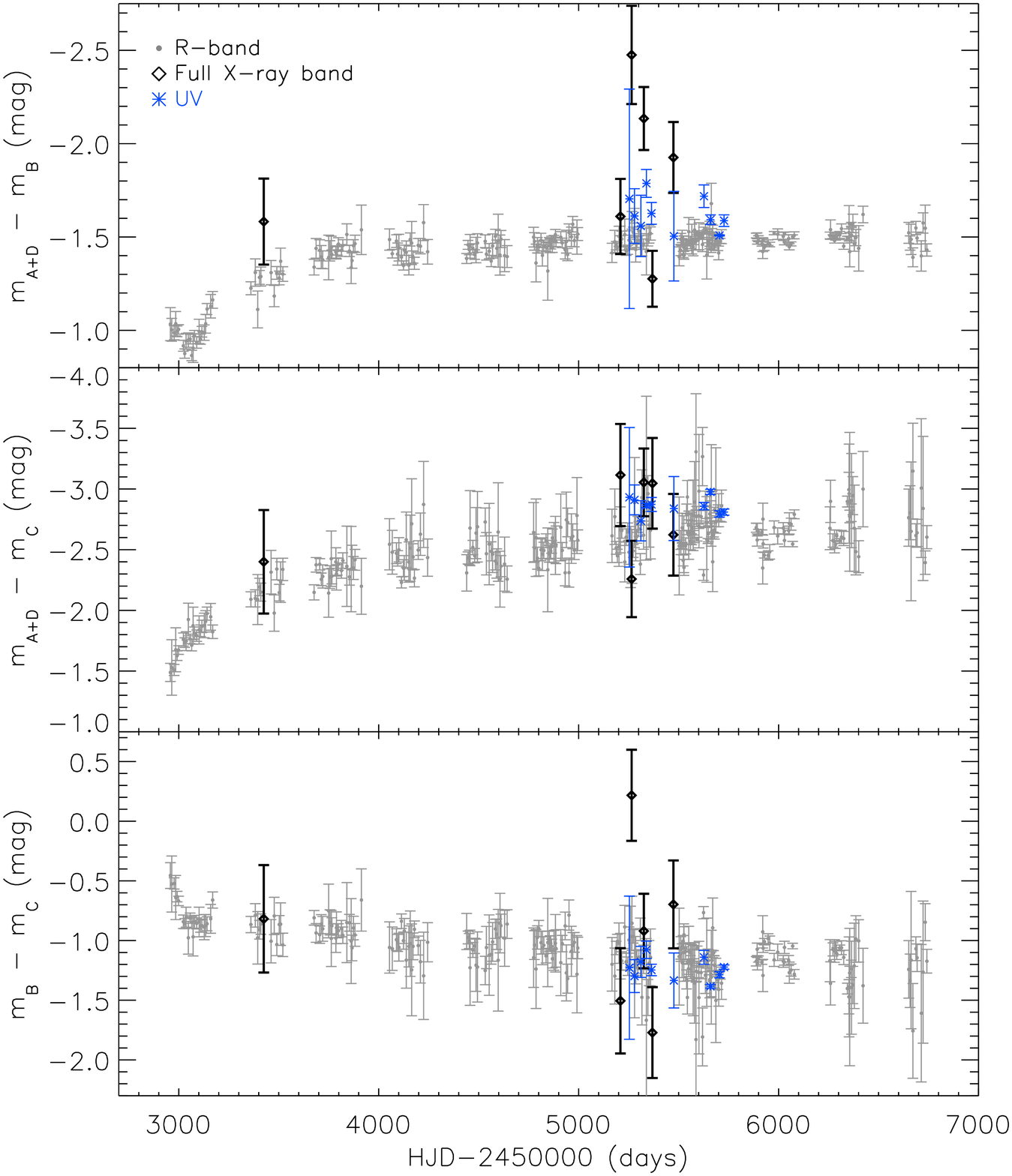}
\caption{\footnotesize Difference light curves in the $R$-band (gray dots), UV (blue asterisks), and full X-ray band (black diamonds)  for SDSS J0924+0219.  Note that while the X-ray and UV fluxes for images A and D are summed here, they are treated as separate in the analysis. }
\label{fig:diffLC}
\end{figure*}

Figure~\ref{fig:diffLC} shows the difference UV and X-ray light curves between (merged) images A+D and B, A+D and C, and between B and C in the top, middle, and bottom panels, respectively. The amplitude of the optical microlensing is clearly largest during the first season. During the seventh and eighth seasons where we have multi-wavelength data, the uncorrelated X-ray variability is pronounced with respect to the optical and UV, while the  uncorrelated UV variability is pronounced with respect to the optical.

%%%%%%%%%%%%%%%%%  MODELS   %%%%%%%%%%%%%%%%%%%%%%%%%%%%%

\section{Analysis}
\label{mod}
We analyze the optical, UV, and X-ray light curves using the technique of \citet{koc04} as adapted for multiwavelength analyses \citep{mor08,mor12,dai10,poi10}. For the present analysis, we are uninterested in the surface brightness profile of the X-ray source, but rather the physical extent, so we exploit the fact that microlensing variability can constrain the half-light radius independent of the surface brightness profile adopted for the source \citep{mor05,con07}. Since SDSS J0924+0219 has negligible time delays ($\lesssim 10$~days) between the images as compared to the microlensing timescales \citep{mor06}, we attribute any residual variability in the difference light curve between two images to microlensing rather than intrinsic quasar variability (unless the timescale for intrinsic variability is shorter than either the time delay or the sampling timescale, which may be important in the X-ray data). Our analysis began with creating a sequence of strong lens models to characterize the macroscopic properties of the lens galaxy. Then, using these lens models, we generated detailed microlensing magnification patterns to represent the distribution of stars in the lens galaxy near each of the quasar images, as described in Section~\ref{sec:micro}. We employed a Monte Carlo simulation to generate many realizations of the model microlensing light curves over a range in source size, using physically-motivated values of the tangential velocity. Finally, we performed a (joint) Bayesian analysis on the acceptable model light curves using the optical, UV, and X-ray monitoring data (Sections~\ref{sec:optonly}--\ref{sec:optx}).

\subsection{Lens Models}
\label{sec:macro}
We started by modeling the strong lensing  in SDSS J0924+0219 to obtain convergence $\kappa$ and shear $\gamma$ estimates in the vicinity of each quasar image.  We modeled the lens galaxy with LENSMODEL\footnote{http://redfive.rutgers.edu/$\sim$keeton/gravlens/} \citep{kee01}, selecting concentric NFW \citep{nav96} and de Vaucouleurs mass profiles to represent the dark matter and stellar mass components, respectively.  Our constraints on the lens astrometry are based on the \emph{HST} $H$-band image from CASTLES, the details of which are in \citet{mor06}. We modeled the lens as an early-type galaxy at redshift $z_l=0.39$ \citep{eig06} with effective radius $r_e=0\farcs31 \pm 0\farcs02$, axis ratio $q = 0.92 \pm 0.02$, and major axis position angle $\theta_e = -27^{\circ} \pm 8^{\circ}$ (east of north). The Einstein ring curve of the quasar host galaxy was included as a constraint on the mass models \citep{mor06,koc01}, and an external shear was included to account for any perturbations from other objects near the lens or along the line of sight. We did not attempt to use the observed flux ratios as constraints on the strong lensing model, since they are anomalous due to microlensing and millilensing from substructure \citep{kee06,eig06,mor06,fau11,bat11}.  

We parametrize the unknown stellar mass fraction $\kappa_*/\kappa$ by creating a sequence of ten models in which the mass of the de Vaucouleurs component is set to the fraction $f_{\rm{M/L}}$ of the mass of the constant mass-to-light model with no dark halo.  We used a uniform sampling from $f_{\rm{M/L}}=1.0$ and 0.1 in steps of 0.1, reoptimizing the lens model for each case. The convergence and shear estimates are provided in Table~\ref{tab:macromodels}.  Finally, by adopting a singular isothermal sphere profile for the lens galaxy, we estimated the stellar velocity dispersion in the lens to be 215~km~s$^{-1}$. Such estimates are known to agree well with dynamical measurements \citep[e.g.,][]{bol08}.

\begin{deluxetable}{c cccc|cccc|cccc}
\tabletypesize{\footnotesize}
\tablewidth{0pt}
\tablecaption{Macroscopic Lens Mass Models}
\tablehead{$f_{M/L}$
                &\multicolumn{4}{c}{Convergence $\kappa$}
                &\multicolumn{4}{c}{Shear $\gamma$}
                &\multicolumn{4}{c}{$\kappa_{*}/\kappa$}\\
		\colhead{}
		&\colhead{A}
                &\colhead{B}
                &\colhead{C}
                &\colhead{D}
                &\colhead{A}
                &\colhead{B}
		&\colhead{C}
		&\colhead{D}
		&\colhead{A}
                &\colhead{B}
                &\colhead{C}
                &\colhead{D}	
                }

\startdata
0.1 & 0.73 &  0.72 &  0.77 &  0.74 &  0.24 &  0.21 &  0.29 &  0.29 &  0.019 &  0.017 &  0.023 &  0.020\\
0.2 & 0.67 &  0.65 &  0.70 &  0.67 &  0.30 &  0.26 &  0.37 &  0.37 &  0.041 &  0.038 &  0.048 &  0.041\\
0.3 & 0.59 &  0.58 &  0.64 &  0.60 &  0.36 &  0.31 &  0.45 &  0.44 &  0.065 &  0.060 &  0.083 &  0.069\\
0.4 & 0.53 &  0.52 &  0.57 &  0.53 &  0.42 &  0.36 &  0.53 &  0.52 &  0.10  &  0.09  &  0.12  &  0.10 \\
0.5 & 0.46 &  0.45 &  0.51 &  0.47 &  0.48 &  0.40 &  0.62 &  0.60 &  0.14  &  0.13  &  0.17  &  0.14 \\
0.6 & 0.39 &  0.38 &  0.44 &  0.40 &  0.54 &  0.45 &  0.70 &  0.68 &  0.20  &  0.19  &  0.24  &  0.20 \\
0.7 & 0.32 &  0.31 &  0.37 &  0.33 &  0.60 &  0.50 &  0.78 &  0.75 &  0.27  &  0.26  &  0.33  &  0.28 \\
0.8 & 0.26 &  0.25 &  0.31 &  0.27 &  0.65 &  0.54 &  0.86 &  0.83 &  0.38  &  0.37  &  0.45  &  0.40 \\
0.9 & 0.19 &  0.18 &  0.24 &  0.20 &  0.71 &  0.59 &  0.95 &  0.90 &  0.58  &  0.57  &  0.65  &  0.60 \\
1.0 & 0.12 &  0.11 &  0.18 &  0.13 &  0.77 &  0.64 &  1.03 &  0.98 &  1.00  &  1.00  &  1.00  &  1.00 \\
\enddata
\tablecomments{Convergence $\kappa$, shear $\gamma$ and the fraction of the total surface density 
composed of stars $\kappa_{*}/\kappa$ at each image location for the series of 
macroscopic mass models.}
\label{tab:macromodels}
\end{deluxetable}

\subsection{Monte Carlo Microlensing Analysis}
\label{sec:micro}

For each of the 10 macromodels parametrized by $f_{\rm{M/L}}$, we generated 40 unique sets of periodic microlensing magnification patterns for each image using the P$^3$M method described in \citet{koc04}. The magnification patterns are 8192 $\times$ 8192 pixel arrays with an outer scale radius of $20 R_E$, where $R_E = 5.7 \times 10^{16} \langle M/M_{\odot}\rangle^{1/2}$~cm is the Einstein radius of a star of mass $M$ in the source plane, yielding a source plane pixel scale of $1.4 \times 10^{14} \langle M/M_{\odot}\rangle^{1/2}$~cm. For a black hole with mass $M_{BH} = 2.8\times 10^8 M_{\odot}$ (as estimated from the MgII emission line width in Morgan et al.\ 2006), this pixel scale is 3.4 times the estimated size of the black hole, $r_g=GM_{BH}/c^2$. These patterns have twice the resolution of those in \cite{mor06}, which is important for simultaneously resolving the optical and X-ray sources. 

  Our simulations are carried out in Einstein units, where source sizes $\hat{r}_s$ and velocities $\hat{v}_e$ are related to their physical scales by the mass of an average lens galaxy star ($r_s=\hat{r}_s\langle M/M_{\odot}\rangle^{1/2}$; $v_e=\hat{v}_e\langle M/M_{\odot}\rangle^{1/2}$).
To convert from Einstein units to physical units, we convolved the probability density for the scaled source size $P(\hat{r}_s)$  with the probability density for the scaled effective velocity,  $P(\hat{v}_e)$, and a statistical model (i.e., a prior) for the true effective source velocity, $P(v_e)$. We constructed $P(v_e)$ using the method described in \citet{koc04}, adopting the peculiar velocity estimates for the redshifts of SDSS J0924+0219 and the lens galaxy from the models presented in \cite{mos11b}. Our velocity model includes the 176~km~s$^{-1}$ projected velocity of the CMB dipole onto the plane of the lens, a probability distribution for the one-dimensional peculiar velocity dispersion of galaxies at $z_l$ with an rms of 272~km~s$^{-1}$, and a one-dimensional stellar velocity dispersion in the lens galaxy of 215~km~s$^{-1}$, based on the lens models. An alternate method of converting to physical units is to assume a prior on $\langle M\rangle$.   In Section~\ref{sec:optonly}, we describe  the optical-only analysis using such a prior.  The joint optical--UV analysis is presented in Section~\ref{sec:optuv}, and  the joint optical--X-ray analysis is presented in Section~\ref{sec:optx}.

\subsubsection{Optical Microlensing Models}
\label{sec:optonly}
The first step in our dual Monte Carlo analysis was to generate $4\times 10^4$ trial light curves for each magnification pattern. We performed 40 separate Monte Carlo simulations of each magnification pattern, attempting a grand total of $1.6\times 10^7$ trials for each source size and image. Trials were made for a grid of 15 different source sizes spanning $13.5 \leq \log{(\hat{r}_{s,{\rm opt}}/{\rm cm})} \leq 17$, yielding $2.4\times 10^8$ total trial fits to the $R$-band data.  We added a systematic error of 0.01 mag in quadrature to the photometric uncertainties for images A+D and B, and a systematic error of 0.025 mag for C.  This adjustment yields a reasonable number of fits with $\chi^2/N_{dof}\approx 1$ and helps compensate for any problems in the photometry such as flux sharing between images due to poor seeing.  Since we did not use the observed flux ratios to constrain our strong lens models, we allowed for 0.5 mag of systematic uncertainty in the absolute flux ratios relative to the macro model.  This helps compensate for any problems in the macro model magnifications and any effects of substructure.  The final goodness of fit, in the sense that the probability of the data given the model is $P(D|M)\propto e^{-\chi^2/2}$, is evaluated from the $\chi^{2}$ of the light curves, where all data have their true statistical weights.  While a model microlensing light curve is generated for each of the four images, the summed model light curve for images A and D is compared to the summed data for these images when computing the $\chi^2$. We dropped all fits with $\chi ^{2}/ N_{\mathrm{dof}\,}> 3.0$, as they contribute negligibly to the Bayesian integrals, leaving a total of $2.8\times 10^7$ complete fits to the optical light curves.

In the process of fitting the microlensing variability, we also identified 18 outliers in the optical data due to poor photometry. We considered five of the best microlensing fits to the original data in which we have first subtracted our model for the intrinsic variability. For each example fit, we identified all observations which were $\geq 3\sigma$ discrepant from adjacent epochs for at least two of the three quasar images.  These observations had systematically worse seeing than average.  Eighteen observations were flagged as outliers in all five example fits, and we exclude these points from the final analysis. These data are enclosed in parentheses in Table~\ref{tab:LC}. It is important to exclude these outliers from the analysis because they would otherwise dominate the $\chi^2$ of the fits and significantly reduce the number of actual good fits that are retained in the Monte Carlo analysis.  Given the individual trials, we then marginalize over the nuisance parameters using uniform priors for the trajectory starting points, the velocity prior described above, and logarithmic priors on the sizes \citep[for full details, see][]{koc04,poi10}. 

\begin{figure}[ht!]
\centering
\includegraphics[scale=.26,trim=4cm 0cm 0cm 2cm,clip=true]{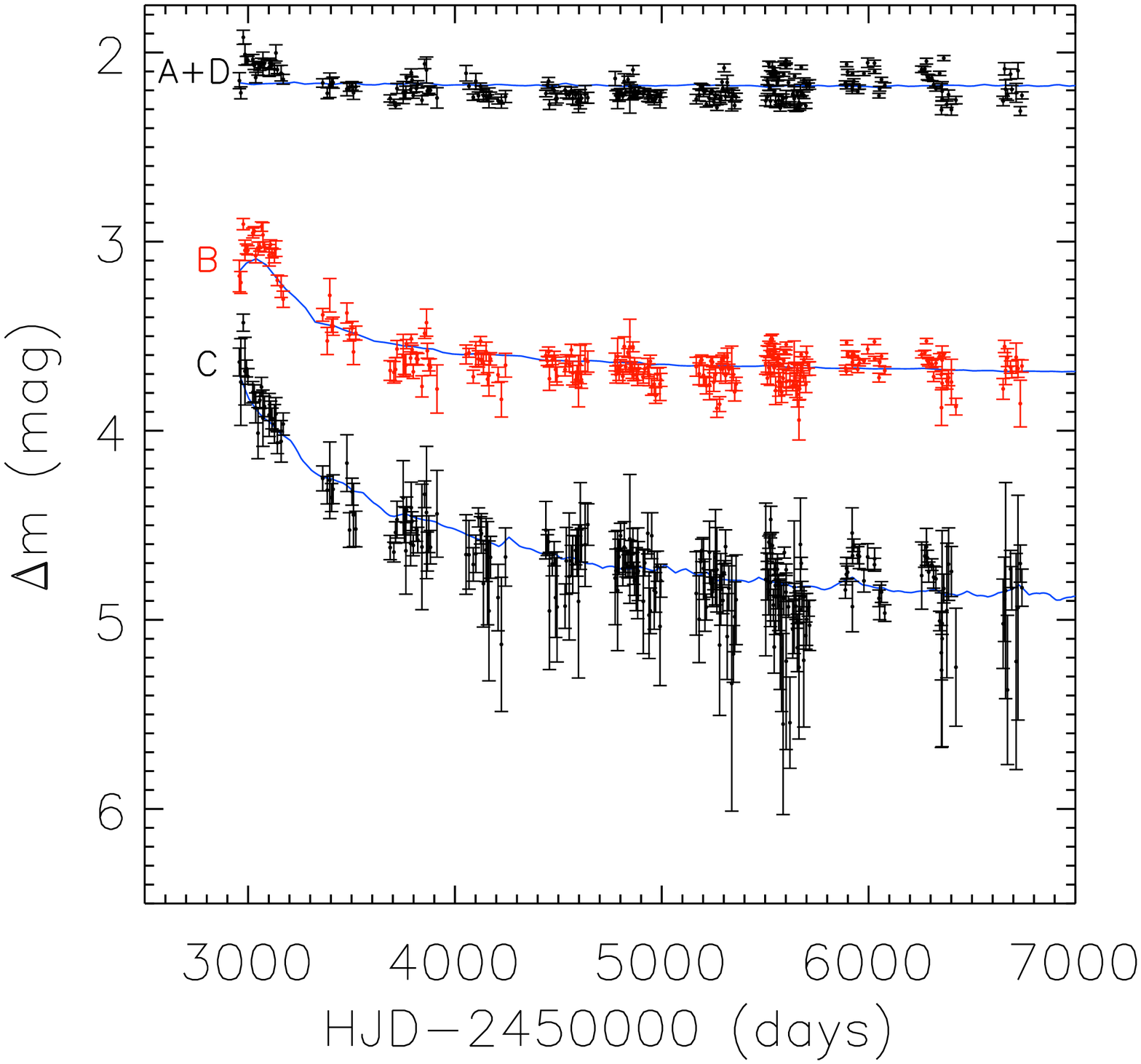}
\includegraphics[scale=.26,trim=4cm 0cm 0cm 2cm,clip=true]{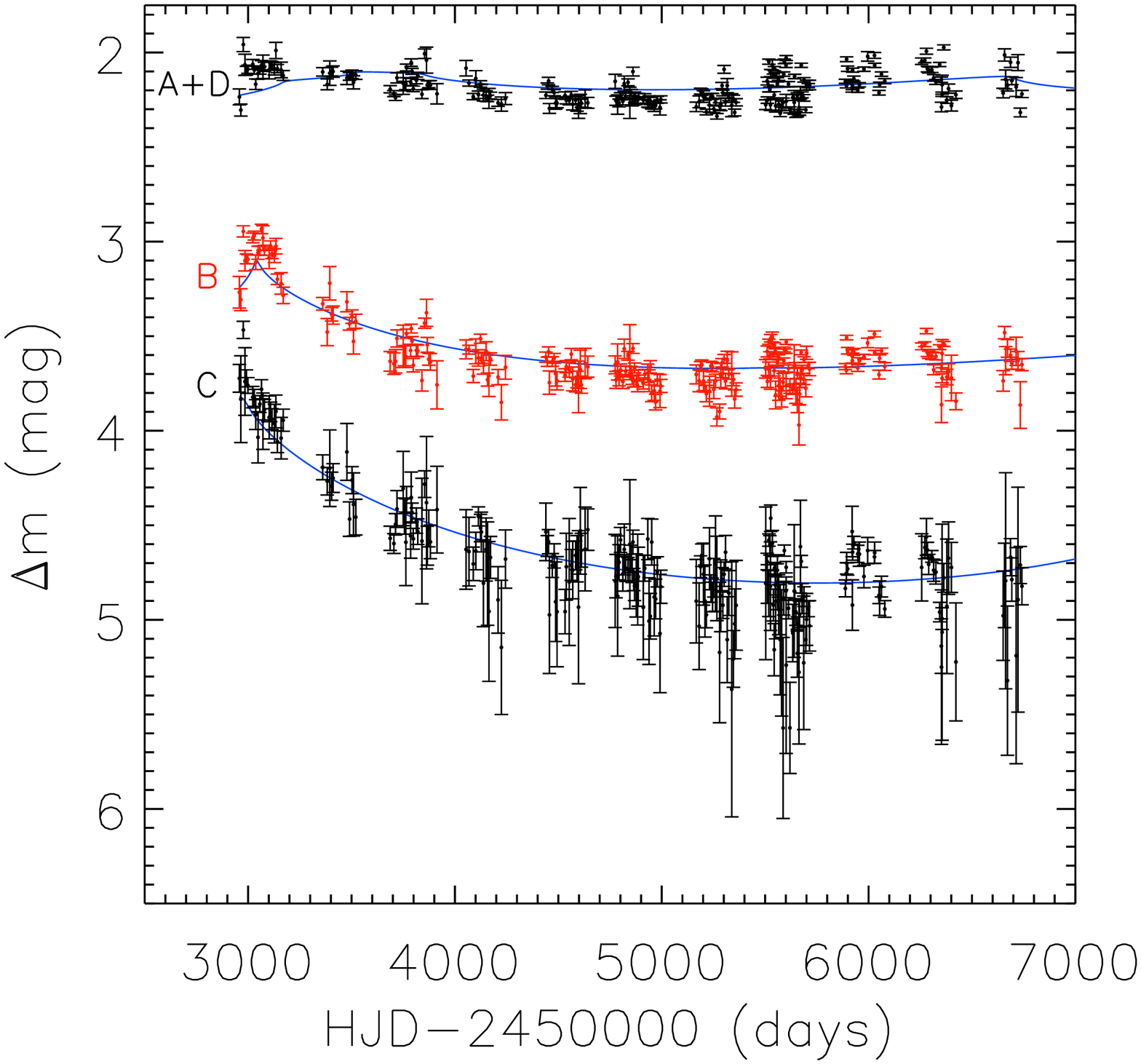}
\caption{\footnotesize  Example optical microlensing model light curves (solid curves) for images A+D, B, and C. The data points with error bars show the data after subtracting the intrinsic source variability. The $\chi^2/N_{dof}$ is 1.1 for both panels.
}
\label{fig:opteg}
\end{figure}

Figure~\ref{fig:opteg} shows example fits to the observed optical flux ratios.  We express the optical source radius as the scale radius $r_{s,{\rm opt}}$ at which the disk temperature corresponds to the rest-frame wavelength of the $R$ band (where $T=h\nu/k$). The cen-

\noindent tral hole in the disk emissivity is neglected to avoid the introduction of an additional parameter, and is small enough to be unimportant. From the optical data alone,  we find a most probable source scale radius of $r_{s,{\rm opt}}=2\times 10^{15}$~cm with a $1\sigma$ confidence interval of $7\times 10^{14}{\rm ~cm}< r_{s,{\rm opt}} < 4\times 10^{15}$~cm. If we impose a uniform prior on the mean microlens mass of $0.1 <\langle M/M_{\odot}\rangle < 1.0$, we obtain  $r_{s,{\rm opt}}=0.9^{+1.4}_{-0.6} \times 10^{15}$~cm. Assuming a typical inclination of $60^{\circ}$ and converting to a half-light radius ($r_{1/2}=2.44 r_s$), we obtain $r_{1/2,{\rm opt}}=6.9^{+7.7}_{-4.5}\times 10^{15}$~cm.  When using a uniform prior on the mean microlens mass as in \citet{mor06}, we obtain  $r_{1/2,{\rm opt}}=3.2^{+4.8}_{-2.3} \times 10^{15}$~cm, consistent with their estimates.

\clearpage
\subsubsection{Optical--UV Joint Analysis}
\label{sec:optuv}

Following \citet{dai10}, all physical variables from the good optical fits (e.g., trajectory and effective transverse velocity) were saved for the combined optical and UV analysis, and fits to the UV light curves were then attempted using the same solution but for 12 different UV source sizes spanning $13\leq \log{(\hat{r}_{s,{\rm UV}}/{\rm cm})} \leq 15.75$.   For the UV data, we added a systematic error of 0.01 mag in quadrature to the photometric uncertainties for each image. This is probably an underestimate of the systematic uncertainties in the UV fluxes (e.g., as created by intrinsic variability), but it allowed us to find reasonable numbers of solutions.  When assessing the quality of our fits to the observed UV flux ratios, we again allowed for 0.5 mag of systematic uncertainty in the flux ratios of the macro models. All fits to the UV data with $\chi^{2}/ N_{\mathrm{dof}}< 30$ were retained (note: the joint $\chi^{2}/ N_{\mathrm{dof}\,}$ remains $<3$). Figure~\ref{fig:xeg} shows an example fit to the observed UV flux ratios.  

The analysis of the combined UV and optical solutions yielded a set of joint probability densities for the variables of interest, namely the UV and optical source sizes. The resulting constraints on $\hat{r}_{s,{\rm UV}}$ and $\hat{r}_{s,{\rm opt}}$ are listed in Table~\ref{tab:sizes}, and the probability distributions are shown in Figure~\ref{fig:plot2}. In physical units, after converting to a half-light radius and correcting both sizes for inclination $i$ assuming $\langle \cos i \rangle=1/2$, the resulting UV size is $10^{14}$~cm~$< r_{1/2,{\rm UV}} < 3\times 10^{15}$~cm, while the optical size is larger at $3\times 10^{15}$~cm~$< r_{1/2,{\rm opt}} < 10^{16}$~cm. Here, we are adopting the results from using a stellar mass prior, as the probability distributions are more stable when altering the $\chi^{2}$ limit on the UV model light curves. The probability distribution for the optical--UV size ratio is shown in Figure~\ref{fig:sratio}.  The probability distribution for $f_{\rm{M/L}}$ resulting from the joint analysis is flat, indicating no preference for the stellar mass fraction, and the best-fit mean stellar mass is $\log{(\langle M/M_{\odot}\rangle)} = -0.5\pm 1.0$.

\subsubsection{Optical--X-ray Joint Analysis}
\label{sec:optx}

Similar to the optical--UV analysis, all physical variables from the good optical fits were saved for the combined optical and X-ray analysis.  The same source sizes used in the UV analysis were attempted for the X-ray fits, and we analyze the microlensing variability in the full, soft, and hard
\begin{figure}[ht!]
\centering
\includegraphics[scale=.2065]{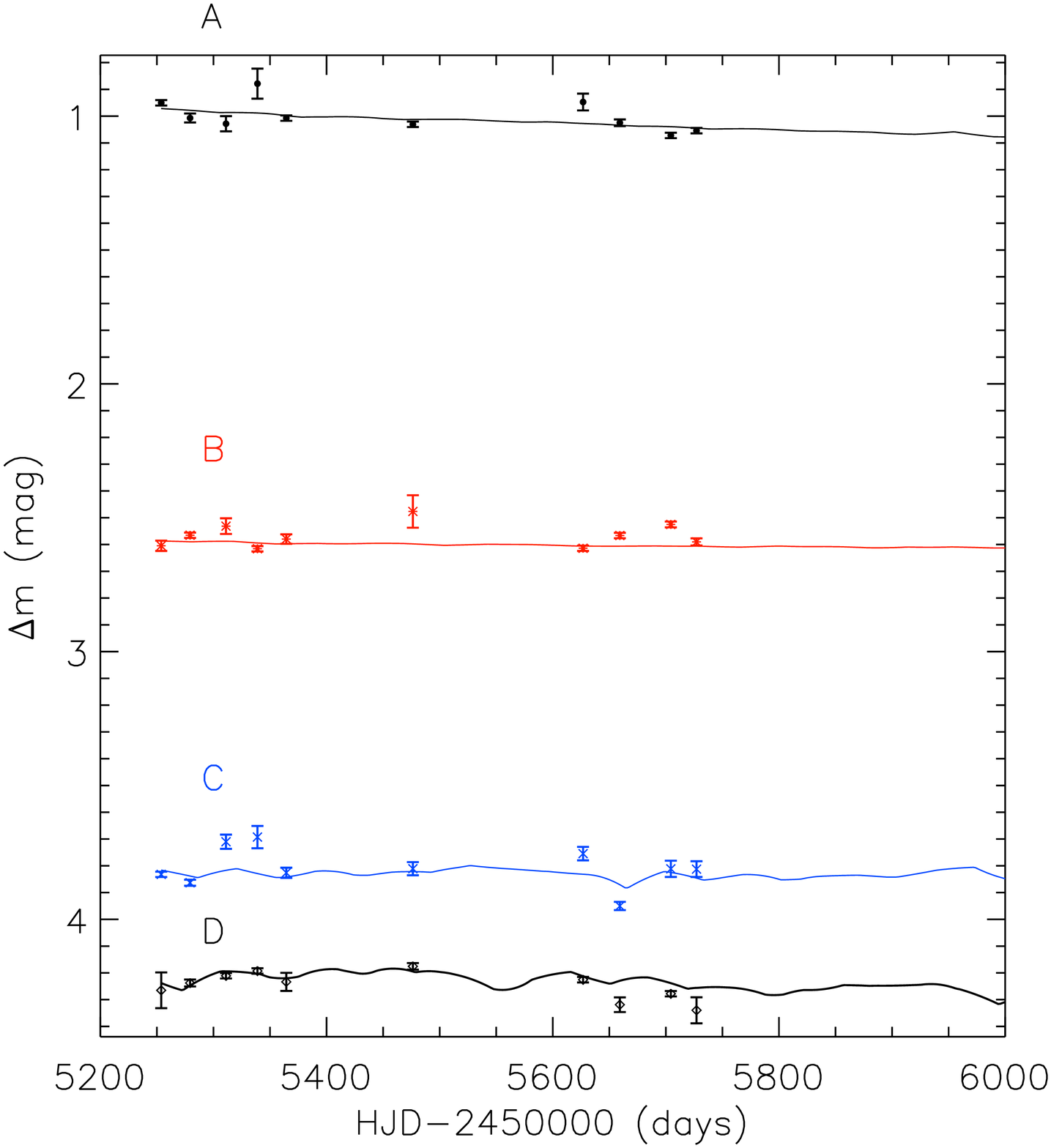}\\
\includegraphics[scale=.2065]{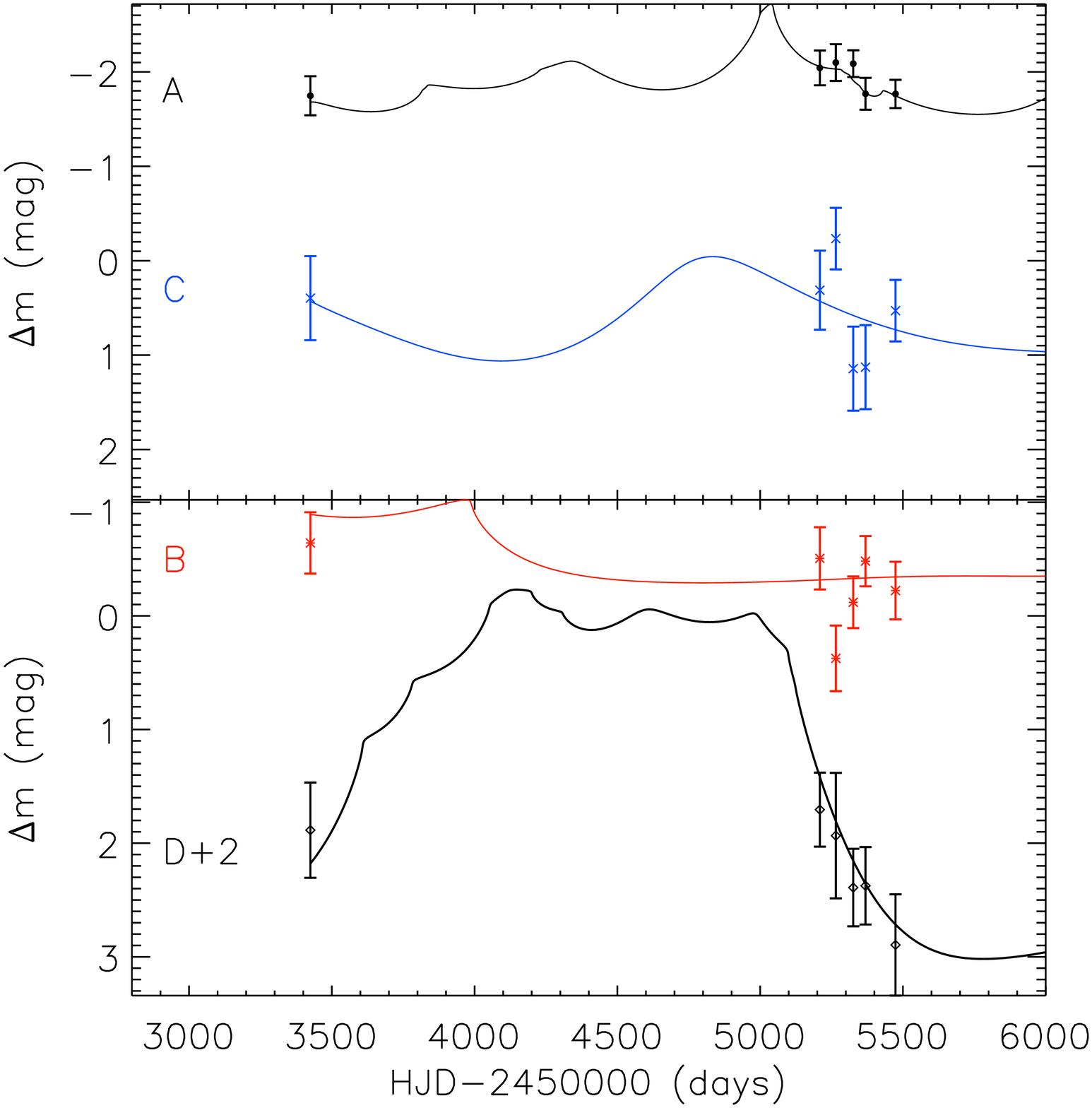}\\
\includegraphics[scale=.2065]{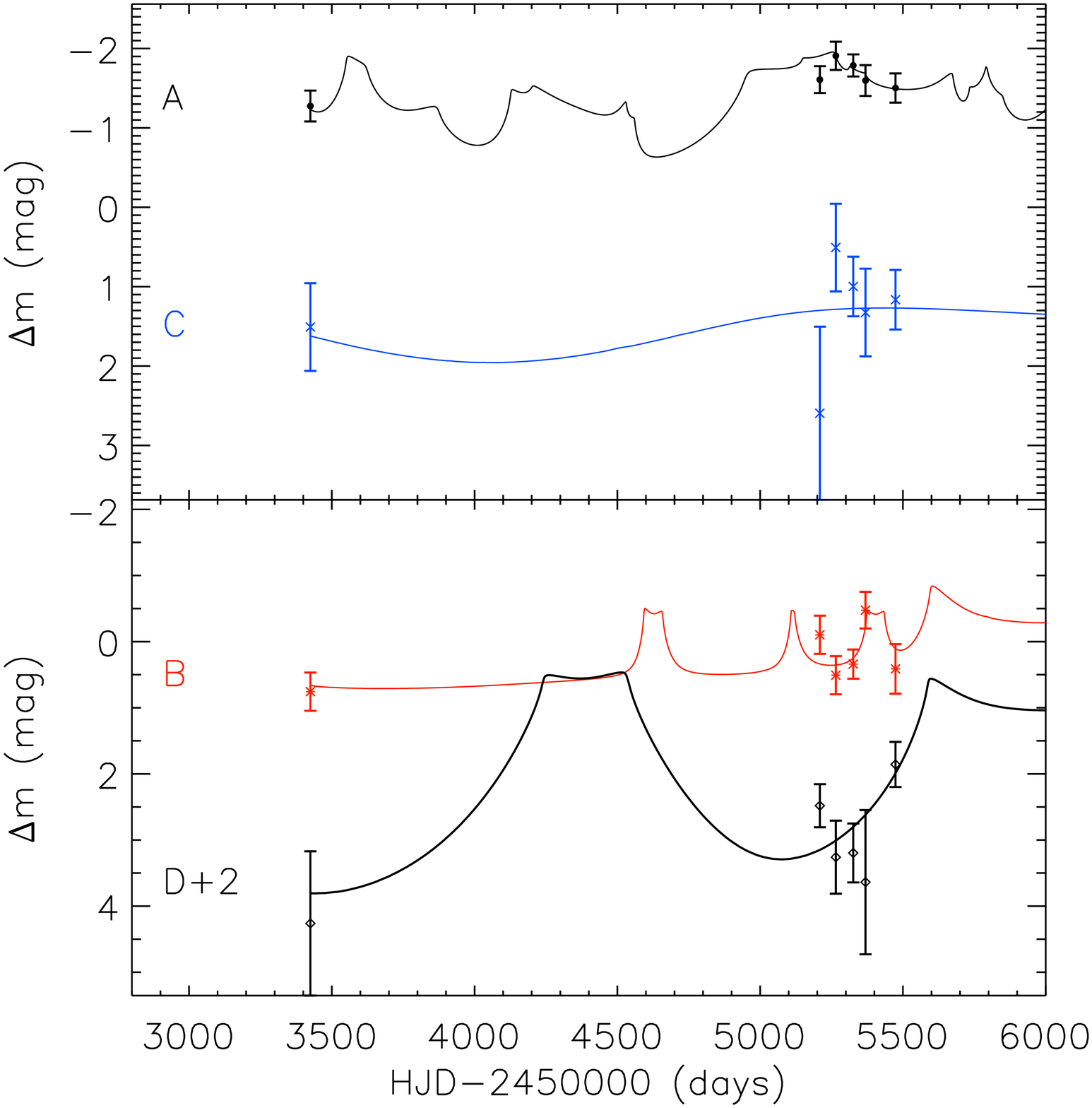}
\caption{\footnotesize Example microlensing model light curves (solid curves) for the UV data (top panel), soft X-rays (middle panel), and hard X-rays (bottom panel). The data points with error bars show the observed data after subtracting our model for the intrinsic source variability.  
}
\label{fig:xeg}
\end{figure}
\clearpage

\begin{figure}[ht!]
\centering
\includegraphics[scale=.28]{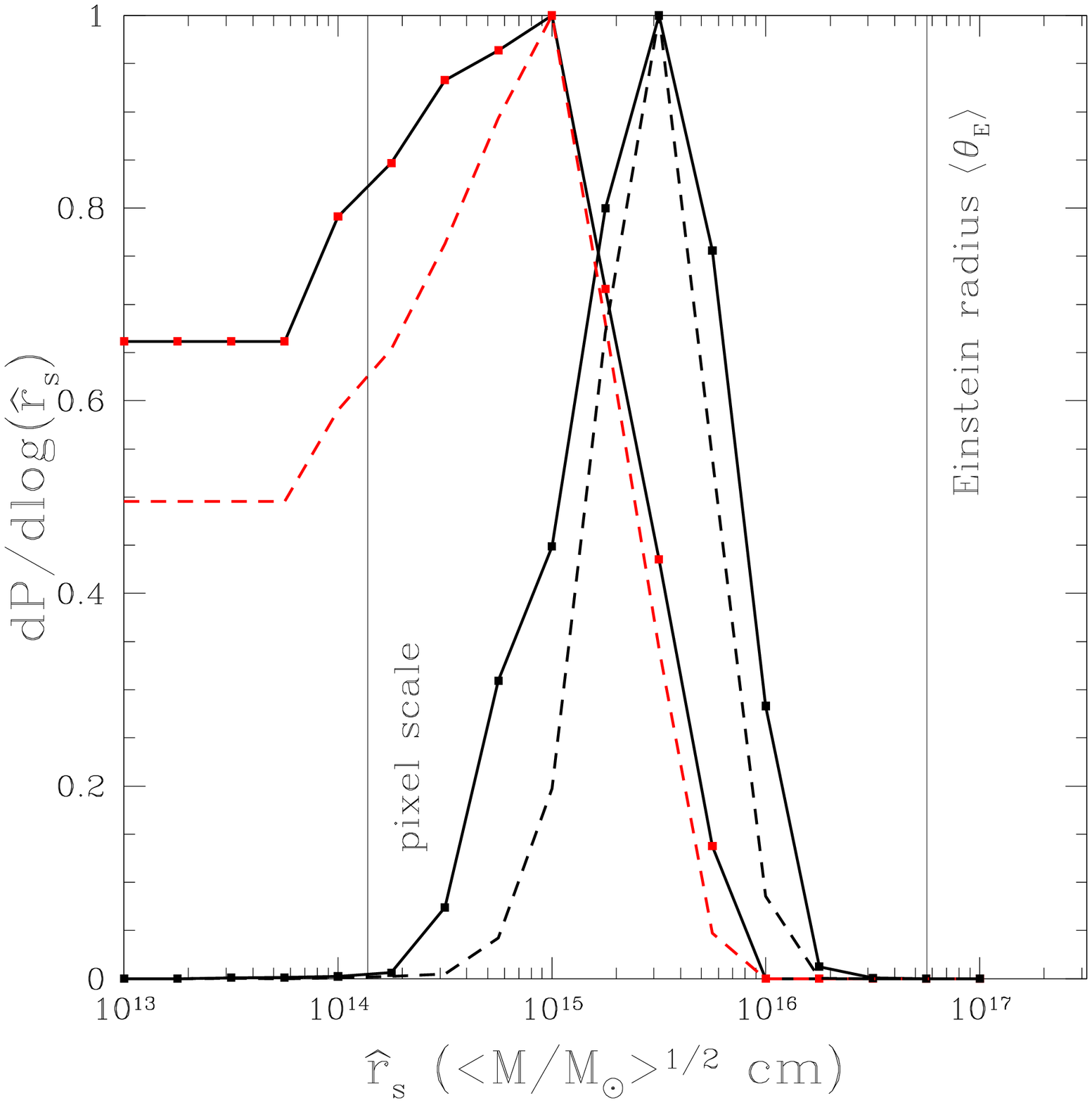}\\
\includegraphics[scale=.28]{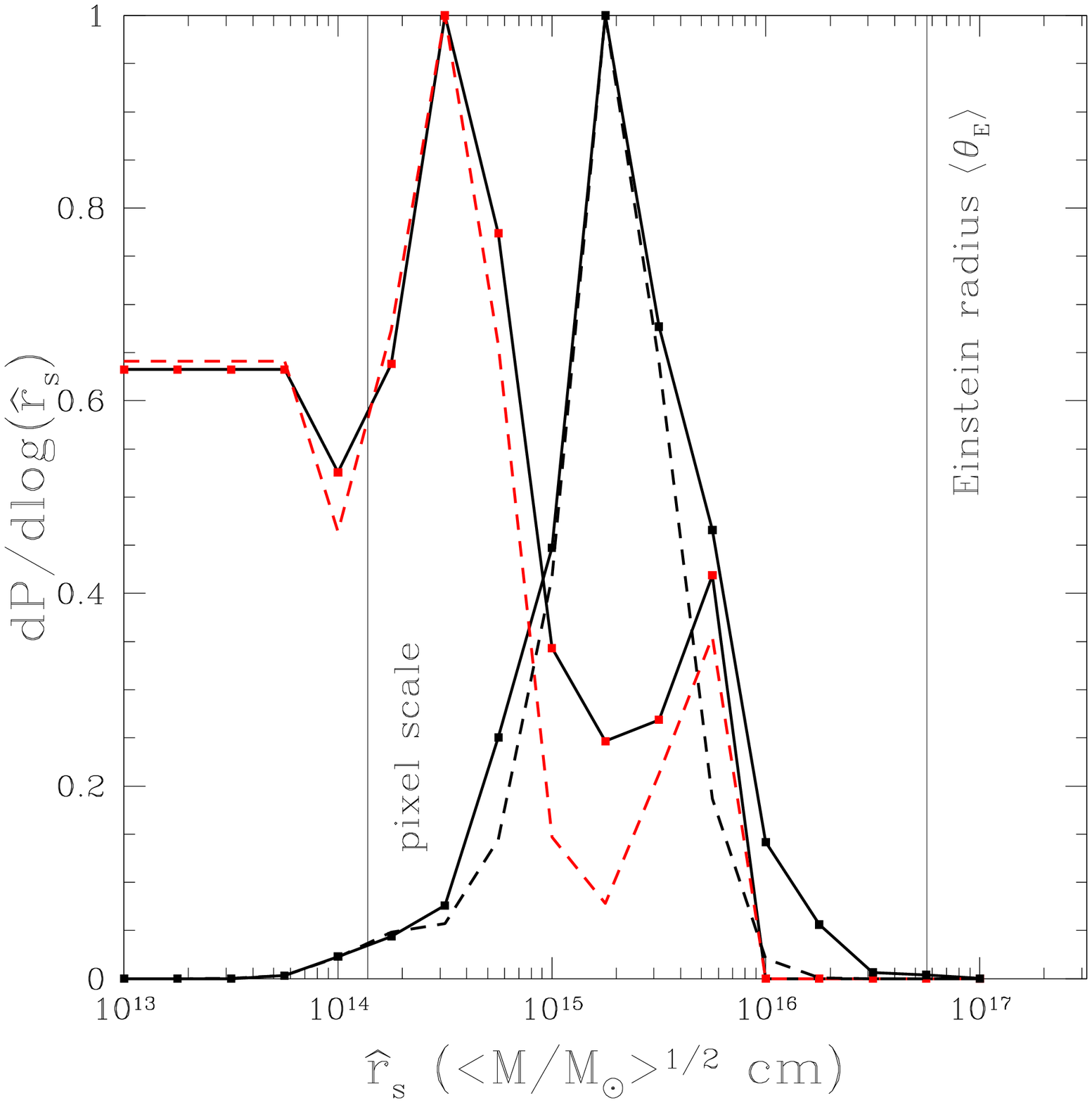}\\
\includegraphics[scale=.28]{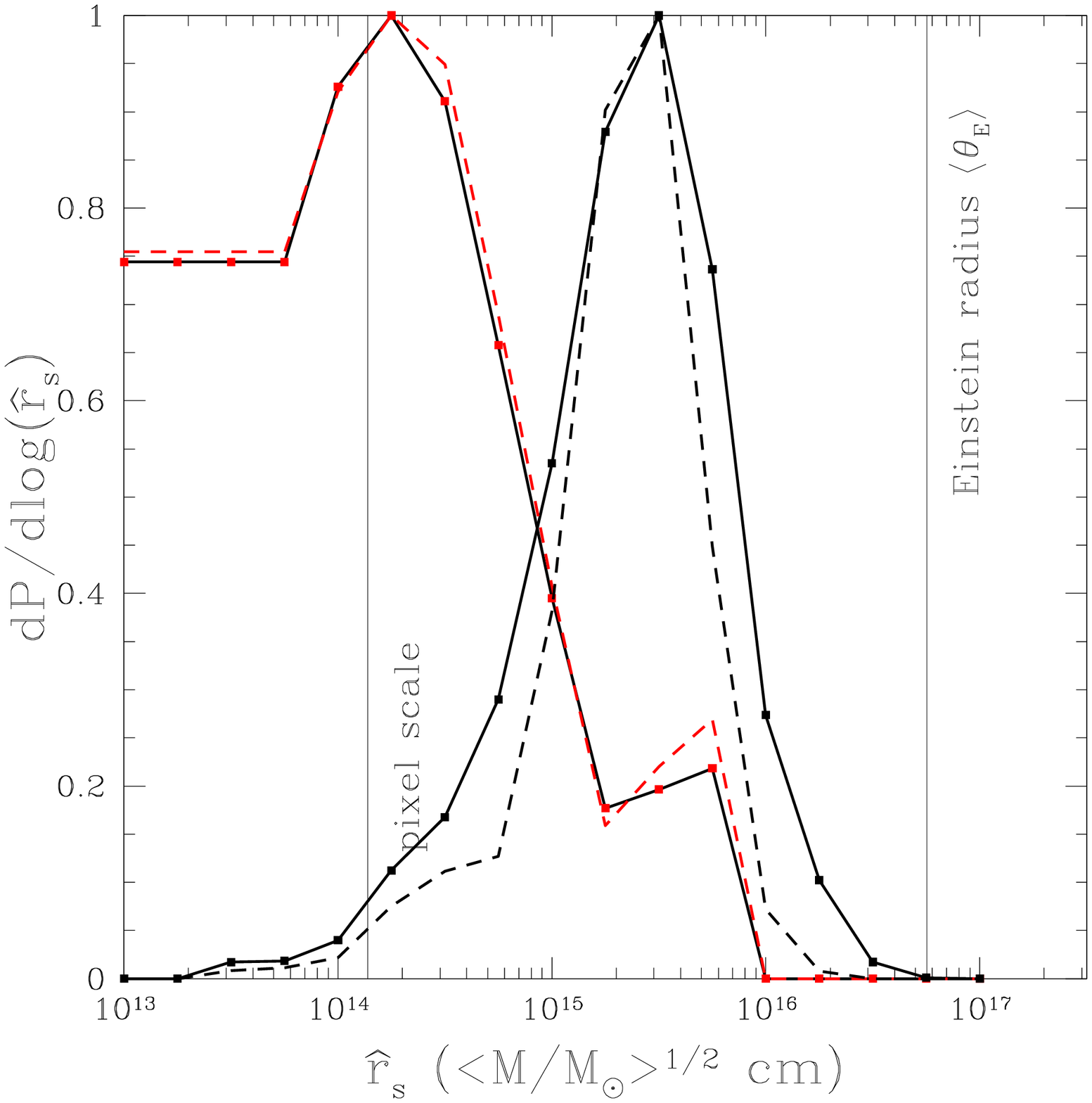}
\caption{\footnotesize 
Probability distributions for the scaled UV  $\hat{r}_{s,{\rm UV}}$ and X-ray $\hat{r}_{s,{\rm X}}$ (red points), and the optical $\hat{r}_{s,\rm opt}$ (black points), from the joint analysis (top: optical--UV, middle: optical--soft-X-ray, bottom: optical--hard-X-ray). The distributions have been scaled to a unit peak height to ease visual comparisons.  The dashed line shows the estimate including a prior of $0.1 < \langle M/M_{\odot}\rangle < 1.0$ on the mass of the stars. The vertical lines show the Einstein radius of the average-mass star and the pixel scale of the magnification map in the source plane.  The expected $r_g$ is at $4.1\times 10^{13} \langle M/M_{\odot}\rangle^{1/2}$~cm.}
\label{fig:plot2}
\end{figure}

\noindent X-ray bands separately.  For the X-ray data, we allowed a systematic uncertainty of 0.1 mag in the photometry for all images,  as this yielded a sufficient number of fits,  and we again allowed a systematic uncertainty of 0.5 mag in the flux ratios.  The former uncertainty should help reduce any contamination from intrinsic variability, which likely has a shorter timescale  than the time delays (6 hrs in the observed frame for X-ray emission within a radius of $\sim 7 r_g$ in SDSS J0924+0219).   All fits with $\chi^{2}/ N_{\mathrm{dof}}< 2.4$, 1.5, and 1.76 for the full, soft, and hard bands, respectively, were retained. Figure~\ref{fig:xeg} shows example X-ray fits from our joint analysis in the soft and hard bands.

The hard-band size falls right at the pixel scale of our magnification patterns, so we can only report an upper limit in this case, while the soft-band size is resolved (Figure~\ref{fig:plot2}). Since the sizes resulting from the joint analysis of the $R$ and soft bands are the most tightly constrained, we quote the best optical size from this analysis.  Hereafter, we report the results from using a stellar mass prior since they are more stable against the chosen cutoff in $\chi^{2}$, as with the UV results.  Table~\ref{tab:sizes} lists the resulting sizes. 

\begin{figure}[ht!]
\centering
\includegraphics[scale=.35]{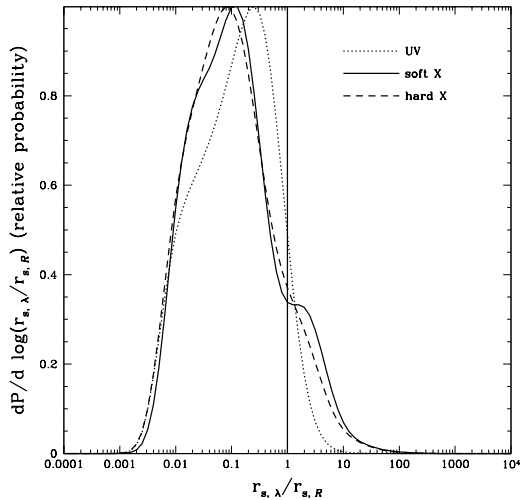}
\caption{\footnotesize 
Probability distributions for the ratios of source sizes (dotted: optical--UV, solid: optical--soft X-ray, dashed:  optical--hard X-ray), scaled to a unit peak height to ease visual comparisons. No sizes have been corrected for inclination.  For a given ratio, the two sizes do not necessarily result from the same trial trajectory. Therefore, these distributions do not take into account the covariances between the source sizes and may be broader than the joint probability distributions. 
}
\label{fig:sratio}
\end{figure}

\begin{figure*}[ht!]
\centering
\includegraphics[scale=.6]{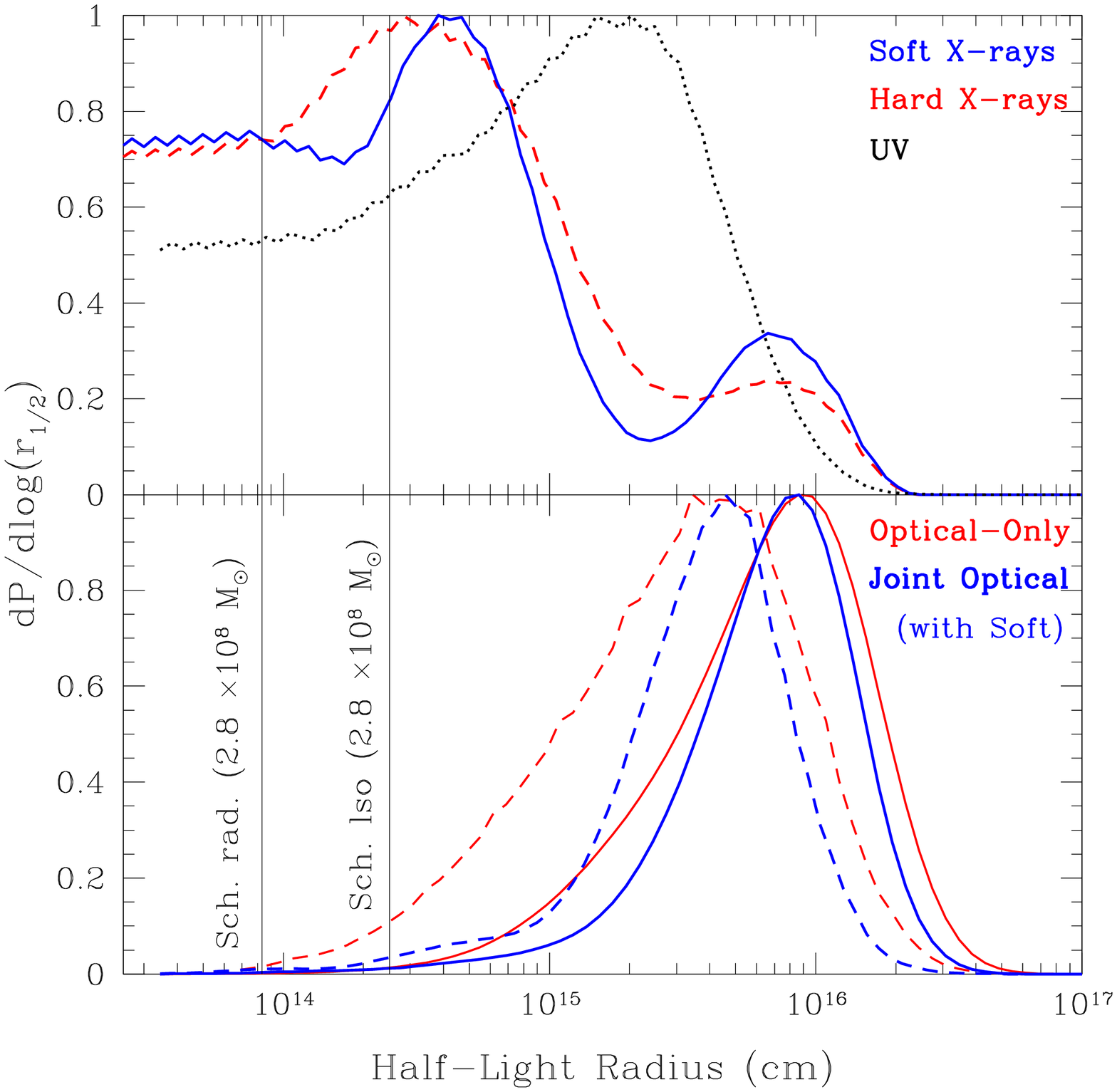}
\caption{\footnotesize Probability densities for the half-light radii of the optical, UV, and X-ray emitting regions, scaled to a unit peak height for ease of visual comparison.  In the top panel, the soft X-ray, hard X-ray, and UV probability densities are shown as a solid blue curve, a dashed red curve, and a dotted black curve, respectively. In the bottom panel, the distributions from our optical-only microlensing analysis are shown in red and are broader than those resulting from our joint (soft-band) analysis, shown in blue. The dashed curves in the bottom panel show the result of imposing a uniform prior on the mean microlens mass of $0.1 < M/M_{\odot}< 1.0$.  All optical and UV sizes have been corrected for a disk inclination of $60^{\circ}$.  The vertical line shows the Schwarzschild radius $R_{BH} = 2GM_{BH}/c^2$ of a $2.8\times 10^8 M_{\odot}$ black hole \citep{mor06}. The last stable orbit for a Schwarzschild black hole is at $3R_{BH}$. The small artifacts in the curves result from coarse binning in the integration of probabilities.  }
\label{fig:plot5dual}
\end{figure*}

In physical units, after converting to a half-light radius and assuming a mean inclination of $\langle \cos i \rangle=1/2$, the resulting optical size is $\sim$1$\sigma$ larger than the previously reported result in \cite{mor06}. The hard-band analysis results in a 1$\sigma$ upper limit of  $r_{1/2,{\rm hard}} < 10^{15}$~cm (or  $r_{1/2,{\rm hard}} < 6\times 10^{15}$~cm at 95\% confidence). Compared to the optical size, the best-fit soft-band size ($3\times 10^{14}$~cm) is 13 times smaller.  The probability distributions for the ratios of the soft- and hard-band sizes to the optical size are shown in Figure~\ref{fig:sratio}, and the individual probability densities are shown in Figure~\ref{fig:plot5dual}. 

\begin{figure}[t!]
\centering
\includegraphics[scale=.4]{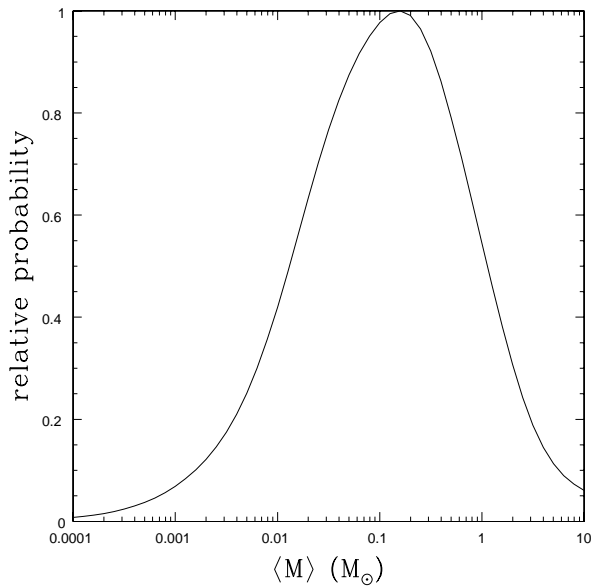}
\caption{\footnotesize  Probability distribution for the mean microlens mass $\langle M\rangle$ for the lens galaxy resulting from our joint optical--soft-X-ray analysis.}
\label{fig:plot4}
\end{figure}

We find a best-fit mean stellar mass of \\ 
$\log{(\langle M/M_{\odot}\rangle)} = -0.9^{+0.8}_{-0.9}$, where  $\langle M\rangle$ is calculated by combining the scaled effective source velocity distribution with the prior probability distribution for the true source velocity. The width of the  $\langle M\rangle$  distribution shown in  Figure~\ref{fig:plot4} reflects the uncertainty in the effective velocity since  $\langle M\rangle \propto \hat{v}_e^{-2}$. Finally, the probability distribution for $f_{\rm{M/L}}$ resulting from our joint analysis is flat, indicating no preference for the stellar mass fraction, in agreement with \cite{mor06}.

\clearpage

\begin{deluxetable}{c c c c c}
\tablecolumns{5}
\tablewidth{0pt}
\tablecaption{Resulting Sizes for SDSS J0924+0219 \label{tab:sizes}}
\tablehead{Analysis & Band & $\log{\hat{r}_s} $ [$\langle M/M_{\odot}\rangle^{1/2}$ cm] & $\log{r_{1/2}}$ [cm] }
\startdata
Opt.--Only&      & $15.2^{+0.5}_{-0.5}  $  &  $15.5^{+0.4}_{-0.6} $  \\
Opt.--UV  & Opt. & $15.4^{+0.4}_{-0.4}  $  &  $15.8^{+0.3}_{-0.3} $  \\
          & UV   & $14.5^{+0.7}_{-1.0}  $  &  $14.9^{+0.6}_{-0.9} $  \\
Opt.--Soft& Opt. & $15.3^{+0.3}_{-0.3}  $  &  $15.6^{+0.3}_{-0.3} $  \\
          & Soft & $14.3^{+0.6}_{-0.8}  $  &  $14.4^{+0.7}_{-0.7} $  \\
Opt.--Hard& Opt. & $15.3^{+0.5}_{-0.3}  $  &  $15.7^{+0.3}_{-0.4} $  \\
          & Hard & $<14.8 $ (68\%)  &  $<15.1 $ (68\%)  \\
Opt.--Full& Opt. & $15.3^{+0.3}_{-0.4}  $  &  $15.4^{+0.3}_{-0.4} $  \\
          & Full & $14.5^{+0.7}_{-1.1}  $  &  $14.8^{+0.6}_{-0.8} $  \\
\tableline
\enddata
\tablecomments{\footnotesize{  All size estimates result from imposing a prior on the mean stellar mass of  $0.1 <\langle M/M_{\odot}\rangle < 1.0$. The optical and UV $r_{1/2}$ estimates have been corrected assuming $\langle \cos i \rangle=1/2$.}}
\end{deluxetable}

\section{Discussion}
\label{disc}

We have improved upon and extended the microlensing analysis of SDSS J0924+0219 by including more recent $R$-band monitoring and more stringent constraints on the microlensing models provided by simultaneous space-based UV and X-ray monitoring. After attempting over 100 million Monte Carlo fits to the light curves, we marginalize over the nuisance variables (trajectory starting points and directions) to generate joint probability densities for the variables of interest ($v_e$, $\langle M\rangle$, $\kappa_*$, $r_s$). 

Adopting the results from the joint analysis, we find a 1$\sigma$ confidence interval for the inclination-corrected optical half-light radius of $1.8\times 10^{15} {\rm cm} <r_{1/2,{\rm opt}} < 7.4\times 10^{15}$, approximately $1\sigma$ larger than the previous measurement of $6.4 \times 10^{14} {\rm cm} < r_{1/2,{\rm opt}} < 2.3 \times 10^{15}$~cm \citep{mor06}. When scaled to a rest wavelength of 2500\AA\ and converted to the scale radius where the photon energy matches the disk temperature  ($T=h\nu/k$), this size corresponds to $\log R_{2500}/{\rm cm}= 15.2^{+0.2}_{-0.3}$.  This size is significantly larger, by 1.5$\sigma$, than the theoretical thin-disk estimate based on the observed, magnification-corrected $HST I$-band flux, similar to the results for other systems \citep{poo07,mor10,bla11,hai13,mos13}. The accretion disk radius based on our optical--only analysis is consistent with but slightly larger than the result in \citet{mor06}, while that based on the joint analysis is shifted to even larger sizes.  Therefore, the combination of using merged A and D images in the $R$-band and including multiwavelength data has likely affected the resulting optical size. Since images A and D could not be reliably separated in the ground-based data, systematic errors may have biased the result in \citet{mor06}, who tried to treat the image fluxes separately. Note that \citet{flo09} found an upper bound on the optical radius that was much larger than the result in \cite{mor06} when adopting the same priors on $\langle M \rangle$.  Aside from merging the images, based on our analysis of the optical, UV, and X-ray emitting regions, the inclusion of UV and X-ray monitoring data seems to provide significant additional information, leading to a more robust constraint on the optical size and further challenging the standard thin disk model.  

SDSS J0924+0219 exhibits pronounced microlensing variability in the UV compared to the optical, indicating a comparatively small extent for the UV continuum emitting region, as expected for an accretion disk. We find a 1$\sigma$ confidence interval of $10^{14}$~cm~$< r_{1/2,{\rm UV}} < 3\times 10^{15}$~cm, which is significantly smaller (by a factor 8) than the optical radius. If we assume that the UV flux originates from the thermal portion of the disk,  our joint optical--UV analysis yields a temperature profile slope of $\beta=0.46\pm 0.46$, where $T\propto r_s^{-\beta}$. While the most probable $\beta$ estimate from our analysis suggests a shallower temperature profile than expected for a standard thin disk, we  cannot rule out the standard thin disk expectation of $\beta=0.75$ given our uncertainties. However, our result is mildly inconsistent with that of \citet{jim14}, who find an average slope of $\beta = 1.33 \pm 0.36$ in a sample of eight lensed quasars. For SDSS J0924+0219, \citet{flo09} obtain $\beta=1.4_{-0.5}^{+1.4}$ based on the anomalous A/D flux ratio in the optical and near-IR bands at a single epoch.   However, this estimate depends critically on the absence of substructure and on the choice of logarithmic or linear priors on the source size. Therefore, we cannot make a clear comparison, as the analysis of \citet{kee06} strongly suggests that at least part of the anomalous A/D flux ratio is not due to microlensing.

The microlensing variability in SDSS J0924+0219 is even more pronounced in the X-rays compared to the optical and UV, indicating an even smaller extent for the X-ray continuum emitting region.  From the soft-band joint analysis, the most probable size ratio is $r_{\rm opt}/r_{\rm soft}=13$.  We were unable to constrain the lower limit for the hard-band source size due to the sparse sampling of the X-ray data and the pixel scale of the magnification patterns, but we find an upper limit of  $r_{1/2,{\rm hard}} < 10^{15}$~cm at 68\% confidence.  These results confirm that the X-ray continuum emitting regions in quasars are compact, supporting models involving an X-ray emitting region concentrated near the inner edge of the accretion disk, along with the results for six other systems \citep[e.g.,][]{mor08,mor12,cha12,mos13,bla13,bla14}. Furthermore, while the three X-ray size measurements are formally consistent, our results suggest that the soft X-ray--emitting region is more extended than the hard X-ray region, although still very compact with respect to the accretion disk.

Note that the optical and UV half-light radii have been corrected assuming an average disk inclination of $\langle \cos i \rangle=1/2$, and the sizes can be rescaled as $r_{1/2}\propto (\cos i)^{-1/2}$. However, other sources of error in physical parameters such as the black hole mass and projected velocity should not have an appreciable effect on our estimated source sizes. The uncertainty in the black hole mass would only affect the size of the central hole in the disk. This would have little effect in the optical and the UV, where the disk model is applicable, and where the effective size is relatively large compared to the last circular orbit. Uncertainties in the projected velocities are fully included and have less effect on the results than naive expectations, as discussed in Kochanek (2004). Lastly, any additional unvarying flux at the location of each image, such as that due to the Einstein ring visible in the \emph{HST H}-band image, will not affect our results, as they use only the time-varying flux ratios between the quasar images. 

 Assuming that microlensing is the sole cause of the flux ratio anomaly, \citet{mor06} predicted that the image A/D magnitude difference ($m_A-m_D$) would increase by at least 0.25 mag over 3 years with a 70\% likelihood.  However,  the flux of image D has remained on average at least a factor 12.5 fainter than that of image A in the optical, and on average a factor $20\pm1$ ($9\pm3$) fainter in the UV (full-band X-rays). Therefore, we still cannot rule out a significant influence from substructure in the lensing potential.  However, since we allow a 0.5~mag uncertainty in the (un-microlensed) flux ratios of the quasar images, our results should not be sensitive to the effects of millilensing. Furthermore, by using a time series analysis, our microlensing models do not rely on precise flux ratio measurements but rather on the time signature of microlensing \citep{koc04,mor12}. Future high-resolution mid-IR observations of this system, such as from the James Webb Space Telescope, should help constrain the effect of substructure in the lens galaxy by providing the intrinsic flux ratios. 

The simulations presented here could be improved by using dynamic magnification patterns (i.e., recalculated for each epoch) as in \cite{poi10}, \cite{mos13}, and \cite{bla14}, although the results should not change significantly. Our results would also be improved by running a joint optical--UV--X-ray analysis, but that is too computationally expensive for our present facilities. Additionally, note that our model for the intrinsic variability is simply based on the correlated variability between the quasar images, and our analysis could be improved by comparing this model to what we expect based on observed light curves for non-lensed quasars \citep[e.g.,][]{mac10}.  Finally, we have neglected the gravitational lensing effects of the black hole which may be important on such small scales, so our simulations may be improved by adopting a fully relativistic disk model \citep[see][]{che13}.  We leave these improvements for a future study.

With further X-ray monitoring for more systems, future studies will be able to more precisely explore the energy structure of the quasar X-ray continuum emission region and test correlations such as the one between black hole mass and X-ray size \citep[e.g.,][]{mos13}.  Continued analysis of the hard and soft X-ray bands should improve our constraints and help distinguish between physical models, such as the X-ray model of \cite{gar14}, and those involving the presence of reflection components (Fabian et al.\ 2004; see review by Uttley et al.\ 2014). For example,  a more extended soft X-ray component would be supported by the physical models of \cite{gar14}, who show that propagation-plus-reflection models fail to explain the soft X-ray lag behavior in the AGN PG 1244+026.  As magneto-hydrodynamical accretion disk models involving an X-ray corona continue to improve \citep[e.g.,][]{nob07,pen10,sch06,sch13,dex14}, so will our understanding of the nature of the various X-ray structures observed through microlensing. With the advent of new survey data such as Pan-STARRS \citep{kai10} and the Dark Energy Survey \citep{hon08}, new lenses will become available for a more extensive analysis of quasar continuum emitting regions.  The Large Synoptic Sky Survey \citep{ive08} is expected to uncover thousands of new lensed quasars, providing well-sampled light curves with multiwavelength microlensing information, which will improve our understanding of the temperature profile of accretion disks and the source of various types of emission in AGN.

\acknowledgments
This material is based upon work supported by the National Science Foundation (NSF) under grant AST-1211146 to CWM. We also gratefully acknowledge support by Chandra X-Ray Center award 1170050.  CSK is supported by NSF grant AST-1009756, and FC and GM are supported by the Swiss National Science Foundation (SNSF).  This research made extensive use of the USNA high performance computing cluster. We thank the reviewer for valuable comments which improved the manuscript. 

{\it Facilities:} \facility{CTIO:2MASS (ANDICAM)}, \facility{HST (NICMOS, ACS)}, \facility{Euler1.2m}, \facility{CXO (ACIS)}

\bibliography{0924micro}
\bibliographystyle{apj}

\end{document}